\documentclass[showpacs,aps]{revtex4}
\emergencystretch=\maxdimen
\usepackage[T1]{fontenc} 
\usepackage{graphics,epsfig}
\usepackage{epstopdf}
\usepackage{graphicx}
\usepackage{dcolumn}
\usepackage[export]{adjustbox}
\usepackage{amsmath,tensor,amssymb}
\usepackage{graphics,epsfig}
\usepackage{epstopdf}
\usepackage{graphicx}

\usepackage[colorlinks=true, citecolor=blue, urlcolor = blue, linkcolor= red, bookmarks=true]{hyperref}

\makeatletter
\def\btt#1{\texttt{\@backslashchar#1}}
\DeclareRobustCommand\bblash{\btt{\@backslashchar}} \makeatother

\begin{document}
	
\title{Phase transition of AdS black holes in 4D EGB gravity coupled to nonlinear electrodynamics}
\author{Sushant G. Ghosh}
\email{sghosh2@jmi.ac.in}
\affiliation{Centre for Theoretical Physics, Jamia Millia Islamia, New Delhi 110 025, India}
\affiliation{Astrophysics and Cosmology Research Unit, School of Mathematics, Statistics and Computer Science, University of KwaZulu-Natal, Private Bag X54001, Durban 4000, South Africa}	
\author{Dharm Veer Singh}
\email{veerdsingh@gmail.com}
\affiliation{Department of Physics, Institute of Applied Science and Humanities, G.L.A University, Mathura, 281406 India.}
\author{Rahul Kumar}\email{rahul.phy3@gmail.com}
\affiliation{Centre for Theoretical Physics, Jamia Millia Islamia, New Delhi 110 025, India}
\author{Sunil D. Maharaj }\email{maharaj@ukzn.ac.za}
\affiliation{Astrophysics and Cosmology Research Unit,	School of Mathematics, Statistics and Computer Science, University of KwaZulu-Natal, Private Bag X54001, Durban 4000, South Africa}

\date{\today}
	
\begin{abstract}
Einstein-Gauss-Bonnet (EGB) gravity is an outcome of quadratic curvature corrections to the Einstein-Hilbert gravity action in the form of a Gauss-Bonnet (GB) term in $ D > 4$ dimensions and EGB gravity is topologically invariant in $4D$. Recently several ways have been proposed for regularizing, a $ D \to 4 $ limit of EGB, for nontrivial gravitational dynamics in $ 4D $. Motivated by the importance of anti-de Sitter gravity/conformal field theory correspondence (AdS/CFT), we analyze black holes with AdS asymptotic to regularized $4D$ EGB gravity coupled to the nonlinear electrodynamics (NED) field. For a static spherically symmetric \textit{ansatz} the field equations are solved exactly,  using two different approaches,  for a NED Lagrangian to obtain an identical solution$-$namely NED charged AdS black holes in $4D$ EGB gravity which retains several known solutions. Owing to the  NED charge corrected EGB black holes, the thermodynamic quantities are also modified, and the entropy does not obey the usual area law. We calculate the heat capacity and Helmholtz free energy, in terms of horizon radii, to investigate both local and global thermodynamic stability of black holes. We observe a secondary Hawking-Page transition between the smaller thermally favoured black hole and thermal AdS space. Our results show that the behaviour of Hawking's evaporation abruptly halts at shorter radii regime such that the black holes do have a thermodynamically stable remnant with vanishing temperature.
\end{abstract}

\maketitle
\flushbottom
	
\section{\label{sec:level1}Introduction}
Lovelock theories \cite{dll,Lovelock:1971yv,Lovelock:1972vz}, with higher-order curvature terms, are generalisations of Einstein's general relativity (GR) to higher dimensions (HD), but the field equations are not of more than second derivatives of the metric functions.  Hence Lovelock theories are free from several problems that affect other higher derivative gravity theories in which the equations of motion are fourth-order or higher, and linear perturbations modes reveal the existence of ghost instabilities.  In the second-order Lovelock theory or Einstein-Gauss-Bonnet (EGB) gravity the action \cite{dll} is supplemented with the quadratic curvature, namely the Gauss-Bonnet (GB)  \cite{Lanczos:1938sf} term apart from the cosmological constant  ($\Lambda$) and the  Ricci scalar ($R$). This special case of EGB gravity has received significant attention for the reason that the EGB action naturally appears in the low energy of heterotic string theory \cite{Gross}. The spherically symmetric static solution in the EGB theory was originally discovered by Boulware and Deser  \cite{bd}, thereby generalising the $D$-dimensional  Scwarzschild-Tangherlini black hole \cite{st}. The Boulware and Deser black hole solution was extended to the charged counterpart by Wiltshire \cite{dw}  and the thermodynamics of black hole was also analysed \cite{dw,ms,egb}.  A series of subsequent interesting works analysed black hole solutions in EGB gravity \cite{egb2} for various sources \cite{hr,Ghosh:2014pga,Ghosh:2014dqa,Lee:2014dha,Graca:2016cbd,Ghosh:2018bxg}, including those coupled to the nonlinear electrodynamics (NED) fields \cite{nedegb,egbnads}, and also in asymptotic AdS spacetime \cite{egbads} due to the  Hawking-Page type transitions. However the EGB  black holes in asymptotically AdS background have the noteworthy characteristic of being thermodynamically favoured for  higher temperatures \cite{egbads,egbnads} as commonly termed Hawking-Page type transitions \cite{hp}. Due to the anti-de Sitter/conformal field theory (AdS/CFT) correspondence, AdS black holes have become more significant to investigate \cite{egbads,egbnads}.  

EGB gravity allows us to explore several conceptual issues in a broader setup; however, in $4D$  the GB term is a topological invariant as its variation is a total derivative with no local dynamics and the theory becomes GR. One requires $D\geq 5$ for nontrivial gravitational dynamics.  Glavan and Lin \cite{gla} proposed a $4D$ EGB gravity by rescaling  the GB  term in $D$-dimensional spacetime,  as $\alpha/(D-4)$ bypasses conditions of Lovelock's theorem \cite{Lovelock:1972vz}.  Further, they considered the limit $D = 4$ at the level of the field equations so that the GB term does make a nontrivial contribution to local dynamics at least for the case of spherical symmetry.  This nontrivial theory will be referred to as the $4D$ EGB theory, which admits spherically symmetric black hole solutions \cite{gla}, generalising the Schwarzschild black holes, and has the repulsive nature of gravity at short distances.  We refer to the process of obtaining a nontrivial $4D$ EGB gravity as \textit{regularization}, which was originally considered by Tomozawa \cite{Tomozawa:2011gp}  with finite one-loop quantum corrections to Einstein gravity, and he also found the spherically symmetric black hole solution. Later this was also done by Cognola {\it et al.} \cite{Cognola:2013fva}  within a classical Lagrangian approach.  Incidentally the identical spherically symmetric  black hole solutions \cite{gla,Tomozawa:2011gp,Cognola:2013fva} have also been obtained in the semiclassical Einstein equations with conformal anomaly \cite{Cai:2009ua} and also in the $4D$ non-relativistic Horava-Lifshitz theory of gravity \cite{Kehagias:2009is}. 

After the \textit{regularization} of $4D$ EGB theory was proposed by Glavan and Lin \cite{gla}, interesting measures have been taken  to investigate the $4D$ EGB gravity \cite{4degb}, which includes generalizing  the black hole solutions \cite{Fernandes:2020rpa,Singh:2020nwo}, Vaidya-like radiating black holes \cite{Ghosh:2020vpc,Ghosh:2020syx}, black holes coupled with magnetic charge \cite{Singh:2020xju,Kumar:2020uyz,Kumar:2020bqf,Kumar:2020xvu}, to the axially symmetric or rotating case (Kerr-like) was also addressed \cite{Wei:2020ght,Kumar:2020owy}, derivation of regularized field equations \cite{Fernandes:2020nbq},  Morris-Thorne-like wormholes \cite{Jusufi:2020yus}, accretion disk around black holes \cite{Liu:2020vkh}, thermodynamics \cite{HosseiniMansoori:2020yfj,EslamPanah:2020hoj,Wei:2020poh},  gravitational lensing  by a black hole \cite{Islam:2020xmy,Jin:2020emq,Heydari-Fard:2020sib,Kumar:2020sag}, and generalization to more general Lovelock gravity theories \cite{Konoplya:2020qqh}. 

Motivated by the above arguments and extensive importance of AdS/CFT correspondence, the aim of this paper is to consider static spherically symmetric black hole solutions with AdS asymptotic to regularised $4D$ EGB gravity coupled to the NED, i.e., we obtain NED charged $4D$ EGB-AdS black holes. Furthermore, following the Kaluza-Klein-like dimensional reduction procedure, we showed that alternate regularized  $4D$ EGB gravity also leads to the identical static spherically symmetric black hole solution. The metric depends on the mass ($M$), GB coupling constant ($\alpha$), and a  parameter $(k)$ coming from the NED field that measures a potential deviation from the $4D$ EGB black holes \cite{gla,Tomozawa:2011gp,Cognola:2013fva}, and they are encompassed as a special case when the charge is switch off ($k=0$). We also find exact expressions for the thermodynamical quantities associated with NED charged $4D$ EGB-AdS black holes, and also perform both local and global thermodynamic stability analysis. The additional parameter $k$ due to NED  provides a deviation from the $4D$ EGB black holes and our model, unlike other previous models \cite{Kumar:2020uyz,Kumar:2020xvu,Singh:2020xju} with de Sitter core, has an asymptotically Minkowski core. 

The paper is organized as follows: Sect.~\ref{sect2} is devoted to a brief review of the gravitational field equations of regularized EGB gravity minimally coupled with the NED field in the $D\to 4$ limit, and the static spherically symmetric NED charged $4D$ EGB-AdS black hole solution is obtained. In Sect.~\ref{sect3} we get the analytical expressions for various thermodynamic quantities and discuss the effect of NED charge on the stability, phase transitions, and P-V criticality. Finally, in Sect.~\ref{sect5}, we summarize the obtained results.
\section{Action,  field equations and solution}\label{sect2}
Before we start our discussion on  $4D$ EGB gravity coupled to the NED field, it is worthwhile to mention that  the {\it regularization} proposed in  \cite{gla,Cognola:2013fva}, is subject to debate  \cite{Ai:2020peo,Hennigar:2020lsl,Shu:2020cjw,Gurses:2020ofy,Mahapatra:2020rds} and a number of questions have been raised.   In addition, several alternate  {\it regularizations} have also been proposed \cite{Lu:2020iav,Kobayashi:2020wqy,Hennigar:2020lsl,Casalino:2020kbt}.  The \textit{regularization} proposed in Refs.~\cite{Lu:2020iav,Kobayashi:2020wqy} leads to a  well defined special scalar-tensor theory, a member of the family of Horndeski  gravity. Hennigar  {\it et al.} \cite{Hennigar:2020lsl} proposed a well defined $D \to 4$ limit of EGB gravity  generalizing the previous work of  Mann and Ross \cite{Mann}. However, the spherically symmetric $4D$  black hole  
solution obtained in \cite{gla,Cognola:2013fva} still  remains valid in these regularized theories \cite{Lu:2020iav,Hennigar:2020lsl,Casalino:2020kbt}.  Hence one can say that these \textit{regularization} procedures lead to exactly the same  black hole solutions \cite{gla,Tomozawa:2011gp,Cognola:2013fva} at least for the case of $4D$ spherically symmetric spacetimes, but may not be valid beyond spherical symmetry \cite{Hennigar:2015mco}. Thus it turns out that the spherically symmetric solution obtained using any of these  {\it regularization} methods will be the same.  However, for convenience, we shall follow the  {\it regularization}  proposed in  Ref.~\cite{gla}. 
 
The action of EGB gravity, which is motivated by the heterotic string theory  \cite{Lanczos:1938sf,Lovelock:1971yv}, coupled to NED becomes  \cite{egbnads}
\begin{eqnarray}
\mathcal{I} &=&\frac{1}{16\pi}\int d^{D}x\sqrt{-g}\left[ {R}-2\Lambda +\alpha{\cal L_{GB}}  \right]+\mathcal{I}_{NED},
\label{action1}
\end{eqnarray}
which contains the Einstein-Hilbert (EH) action $ R$, a cosmological term,
$\Lambda=-{(D-1)(D-2)}/{2l^2}$, and 
the GB quadratic curvature correction given by 
 \begin{equation}
\mathcal{L}_{GB}=R_{\mu\nu\gamma\delta}R^{\mu \nu\gamma\delta}-4R_{\mu\nu}R^{\mu\nu}+R^{2}.
\end{equation}
The GB coupling constant $\alpha>0$ is of dimension $ [length^2] $, and it is related to the string scale. The matter is  
described by the action $\mathcal{I}_{NED}$, which for the NED reads 
\begin{equation} 
\mathcal{I}_{NED}=-\frac{1}{4\pi}\int d^D x\sqrt{-g}\,\, \mathcal{L}(F),
\end{equation}
where the Lagrangian density $  \mathcal{L}(F) $, is an arbitrary  continuous function of the  invariant $F=\frac{1}{4}F_{ab}F^{ab}$ and 
$F_{ab}=\partial _{a}A_{b}-\partial_{b}A_{b}$, with $A_a$ being the potential for the NED charge.
Varying the action (\ref{action1}), we obtain the equations of motion  \cite{egbnads}
\begin{eqnarray}
&&{G}_{a b} +\alpha {H}_{a b} = \mathcal{T}_{ab}\equiv2\left[\frac{\partial {\mathcal{L}(F)}}{\partial F}F_{a c}\tensor{F}{_b}^{c}-g_{a b}{\mathcal{L}(F)}\right], \label{FieldEq} \\
&& \nabla_{a}\left(\frac{\partial {\mathcal{L}(F)}}{\partial F}F^{a b}\right)=0\qquad \text{and} \qquad \nabla_{a}(^*F^{ab})=0,
\label{egb3}
\end{eqnarray}
where $G_{ab}$ and $H_{ab} $, respectively, are the Einstein tensor and the Lanczos tensor \cite{Lanczos:1938sf}:
\begin{eqnarray}
&&G_{ab}=R_{ab}-\frac{1}{2}g_{ab}R - \Lambda g_{ab},\nonumber\\
&&{H}_{ab}=2\left[RR_{ab}-2R_{a c}R^{c}_{b}-2R^{c d}R_{a c bd} +R_{a}^{~c d e}R_{b c d e}\right]-{1\over 2}g_{ab}{\mathcal{L}}_{GB},
\end{eqnarray}
with energy  momentum tensor \cite{egbnads}
\begin{eqnarray}
&&\mathcal{T}_{ab}=-\frac{2}{\sqrt{-g}}\frac{\delta \mathcal{I}_{NED}}{\delta g^{ab}}, \\
&&\tensor{\mathcal{T}}{^a}{_b}=2\left[\frac{\partial {\mathcal{L}(F)}}{\partial F}F^{ac}F_{b c}-\delta^{a}_{b}{\mathcal{L}(F)}\right].
\end{eqnarray}
We wish to obtain static spherically symmetric black hole solutions of Eq.~(\ref{FieldEq}).  We consider the metric to be of the following form \cite{Ghosh:2014pga}
\begin{equation}
ds^2 = -f(r)dt^2+\frac{1}{f(r)} dr^2 + r^2 d\Omega^2_{D-2},
\label{metric}
\end{equation}
where $d\Omega_{D-2}$ is the metric of a $(D-2)$-dimensional unit sphere and $\mathcal{T}_{a b}$ is the energy momentum tensor of matter that we consider as a NED as in Ref.~\cite{Ghosh:2018bxg}, where the Lagrangian density reads
\begin{equation}
\mathcal{L}(F) =\beta\,F \exp \left[-k q^{-\gamma}(2F)^{\zeta}\right],\label{lag}
\end{equation}
where
$ \beta={(D-2)(D-3)}/{2} $, $\gamma = {(D-3)}/{(D-2)} $, $ \zeta={(D-3)}/{(2D-4)},$ $q$ is magnetic charge, and $k$ is NED parameter. The  Maxwell field reads 
\begin{eqnarray}
F_{\mu\nu}&=&2\delta^{\theta_{1}}_{[\mu}\delta^{\theta_{2}}_{\nu]}q\sin\theta_{1}; \qquad\qquad\qquad\qquad\qquad \qquad\qquad D=4,\nonumber\\
F_{\mu\nu}&=&2\delta^{\theta_{D-3}}_{[\mu}\delta^{\theta_{D-2}}_{\nu]}\frac{q^{D-3}}{r^{D-4}}\sin\theta_{D-3}\left[\prod_{j=1}^{D-4}\sin^2\theta_{j}\right];\qquad  D\geq 5.
\label{ee3}
\end{eqnarray}
Eq. (\ref{egb3}) implies that $dF=0$ so that we obtain
\begin{equation}
q'(r)2\delta^{\theta_{D-3}}_{[\mu}\delta^{\theta_{D-2}}_{\nu]}\frac{q^{D-3}}{r^{D-4}}\sin\theta_{D-3}\left[\prod_{j=1}^{D-4}\sin^2\theta_{j}\right] d\theta\wedge d\phi\wedge\hdots\wedge d\phi_{(D-2)} =0.
\end{equation}
This leads to $q(r)=q=$ constant magnetic charge.  
Hence the field strength tensor $F_{\theta\phi}$ is
\begin{equation}
F_{\theta\phi}=\frac{q(r)}{r^{D-4}}\sin\theta_{D-3}\left[\prod_{j=1}^{D-4}\sin^2\theta_{j}\right],\qquad 
\end{equation}
The other components of $F_{\mu\nu}$ have negligible influence in comparison of $F_{\theta\phi}$. Then $F$ and $\mathcal{L}(F)$ are, respectively, simplified as
\begin{eqnarray}
F=\frac{q^2}{2r^{2(D-2)}}\quad \text{and} \quad \mathcal{L}(F)=\frac{\beta q^2}{2 r^{2(D-2)}}\exp\left[-\frac{k}{r^{D-3}}\right].\label{lf}
\end{eqnarray}
 The energy momentum tensor can be given as
\begin{eqnarray}\label{em}
T^t_t=T^r_r =\frac{(D-2)Mk\; \exp \left( {{-k}/{r^{D-3}}} \right)}{r^{2D-4}} = \rho(r).
\label{emt}
\end{eqnarray}
The Bianchi identity  $T^{ab}{}_{;b} =0$  gives 
\begin{equation}\label{EMT}
0= \partial_r T^r_r + \frac{1}{2} g^{00} \left[T^r_r-T^0_0\right] \partial_r g_{00} 
+ \frac{1}{2} \sum g^{ii}  \left[T^r_r-T^i_i\right] \partial_r g_{ii},
\end{equation}
\begin{eqnarray}\label{EMT1}
T^{\theta}_{\theta} &=& T^{\phi}_{\phi} = T^{\psi}_{\psi} = \rho(r) + \frac{r}{D-2} 
\partial_r \rho_{\theta}(r), 
\end{eqnarray}
and the energy momentum tensor  is completely specified by (\ref{em}) and (\ref{EMT1}) and corresponds 
to an anisotropic fluid. 
We are interested in a static spherically symmetric solution of $4D$ EGB gravity coupled to NED which can be obtained by integrating    the $(r,r)$ equation of (\ref{FieldEq}) in the limit $D\to 4$,  and the solution reads as
\begin{eqnarray}
f_{\pm}(r)=1+\frac{r^2}{2\alpha}\left(1\pm\sqrt{1+4\alpha\left(\frac{2M \exp(-k/r)}{r^3}-\frac{1}{l^2}\right)}\,\right),
\label{sol1}
\end{eqnarray} 
by appropriately relating $M$  with integration  constants. For $D=4$, the Lagrangian density in Eq.~(\ref{lag}) can be expanded as follows
\begin{equation}
\mathcal{L}(F)\approx F - \frac{2^{1/4}k}{\sqrt{q}}F^{5/4} + \frac{k^2}{\sqrt{2}q}F^{3/2} -\frac{k^{3}}{3\times 2^{1/4}q^{3/2}}
F^{7/4} +\frac{k^4}{12q^2}F^2 -\frac{k^5}{30\times 2^{3/4}q^{5/2}} F^{9/4}+\mathcal{O}\left(F^{5/2}\right),
\end{equation}
which in the weak-field limits smoothly goes over to the Maxwell linear electrodynamics, i.e., $\mathcal{L}(F)\to F $, whereas in the strong-field limits it vanishes.
Solution
(\ref{sol1}) is an exact solution of the field equation (\ref{FieldEq}) for stress energy tensor $ \mathcal{T}_{ab} $. The charged $4D$ EGB black hole solution (\ref{sol1}), in absence of the NED field, $k=0$, reduces to the Glavan and Lin \cite{gla} solution (\ref{egb3}), and for $r\gg k$ it becomes
\begin{eqnarray}
f_{\pm}(r)=1+\frac{r^2}{2\alpha}\left(1\pm\sqrt{1+4\alpha\left(\frac{2M }{r^3} - \frac{q^2}{r^4}-\frac{1}{l^2}\right)}\,\right) + \mathcal{O}\left(\frac{1}{r^3}\right).
\label{sol2}
\end{eqnarray}
We note that the above form is exactly that of the Maxwell charged black hole of the  $4D$ EGB gravity with a negative cosmological constant \cite{Fernandes:2020rpa},  if one identifies the electric charge as $q^2 = 2Mk$ \cite{Ghosh:2018bxg}. The analogous Hayward \cite{Kumar:2020xvu} and Bardeen \cite{Kumar:2020uyz,Singh:2020xju} black holes in $4D$ EGB gravity do not go over to Maxwell charged black holes. Whereas the Born-Infeld $4D$ EGB black hole is sourced by the electric field \cite{Yang:2020jno}, contrary to magnetically-charged black hole (\ref{sol1}).  The  advantage of nonsingular black hole solutions  (\ref{sol1}), the exponential mass function arising due to NED leads to a Minkowski-flat core around $ r=0 $, which is in striking contrary with other regular black holes that generally have de-Sitter core \cite{Kumar:2020uyz,Kumar:2020xvu,Ghosh:2018bxg,Kumar:2019pjp,Singh:2020xju}.
Whereas, in the limit $\alpha \to 0$ or large $r$,  the  solution (\ref{sol1})  behaves asymptotically as 
\begin{eqnarray}\label{solGR}
&&f_-(r) \approx 1-\frac{2M}{r}\exp\left[-\frac{k}{r}\right]+ \frac{r^2}{l^2} + \mathcal{O}\left(\frac{1}{r^3}\right),\nonumber\\
&&f_+({r})\approx 1+\frac{2M}{r}\exp\left[-\frac{k}{r}\right]-\frac{r^2}{l^2}+\frac{r^2}{\alpha}+ \mathcal{O}\left(\frac{1}{r^3}\right),
\end{eqnarray}
the $-$ve branch corresponds to the $4D$ regular AdS black hole \cite{Culetu:2015cna}, whereas the $+$ve branch does not lead to a physically meaningful solution as the positive sign in the mass term indicates instabilities of the graviton \cite{bd}, and hence we shall confine ourselves to the $-$ve branch of the solution (\ref{sol1}).   Further if we take the limit $r\rightarrow\infty$ or $M=0$ in solution (\ref{sol1}),  the  $-$ve branch of the solution (\ref{sol1}) is asymptotically flat whereas the $+$ve branch of the solution (\ref{sol1}) is asymptotically dS (AdS) depending on the sign of $\alpha$ $(\pm)$.    
It is known that most regular or nonsingular black holes coupled to NED have a core that is asymptotical de Sitter (with constant positive curvature) \cite{Ghosh:2018bxg,Singh:2020xju,Kumar:2019pjp,Kumar:2020xvu,Kumar:2020uyz,Bardeen1,AyonBeato:1998ub,Hayward:2005gi,Dymnikova:1992ux,Fan:2016hvf}. However, the regular black hole described by the metric (\ref{sol1})  has an asymptotically Minkowski core \cite{Simpson:2019mud} thereby greatly simplifying the physics in the deep core. Further, the GR branch (\ref{solGR}) of  (\ref{sol1}) reproduces exactly Schwarzschild de Sitter solution in the absence of NED ($k=0$).

These models (\ref{sol1}) with exponential function greatly simplify the physics in the deep core \cite{Simpson:2019mud}. Also, one can get rid off rather a messy cubic, and quartic polynomial equations arise in Bardeen \cite{Kumar:2020uyz} and Haywards  \cite{Kumar:2020xvu}	and finally, this model is mathematically interesting due to its tractability, and physically attractive, unlike de Sitter core, having Minkowski core asymptotically \cite{Simpson:2019mud}. In what follows, we shall show that the our black hole (\ref{sol1}) shares some similar features like other regular black holes, but there are also striking differences.

\subsection{Horizons and extremality }
\begin{figure*} [h]
	\begin{tabular}{c c c c}
		\includegraphics[width=0.52\linewidth]{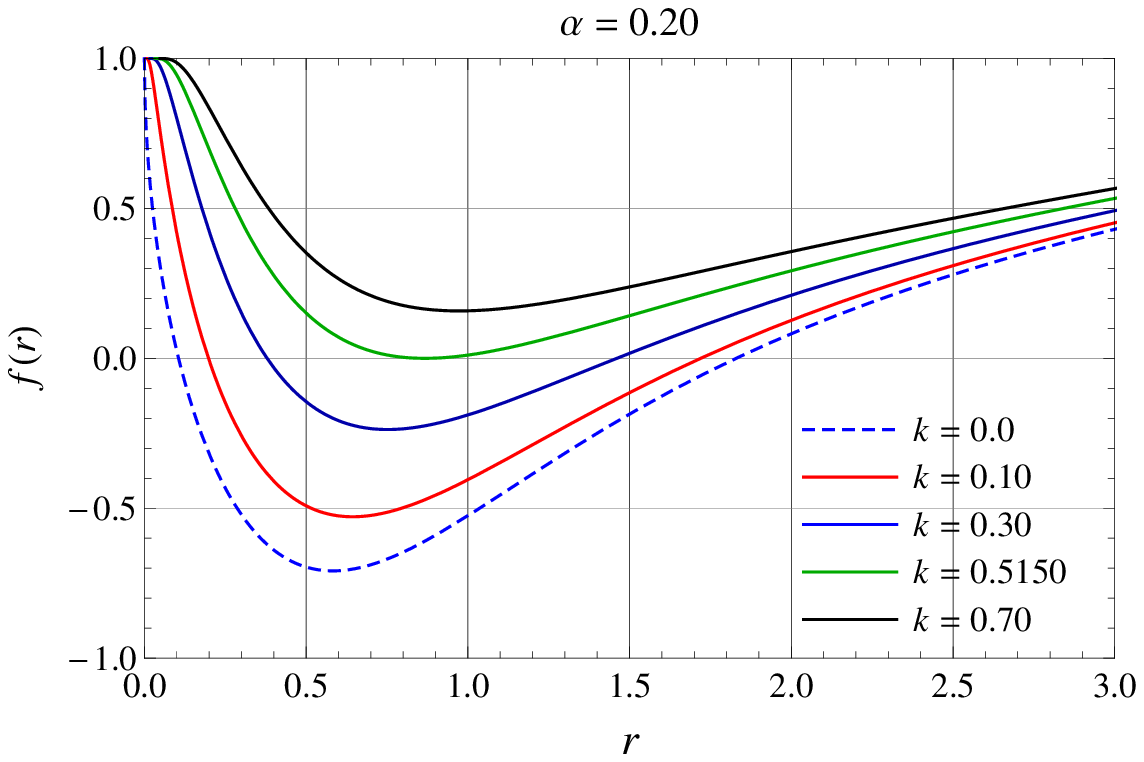}
\hspace*{-0.5cm}		 
        \includegraphics[width=0.52\linewidth]{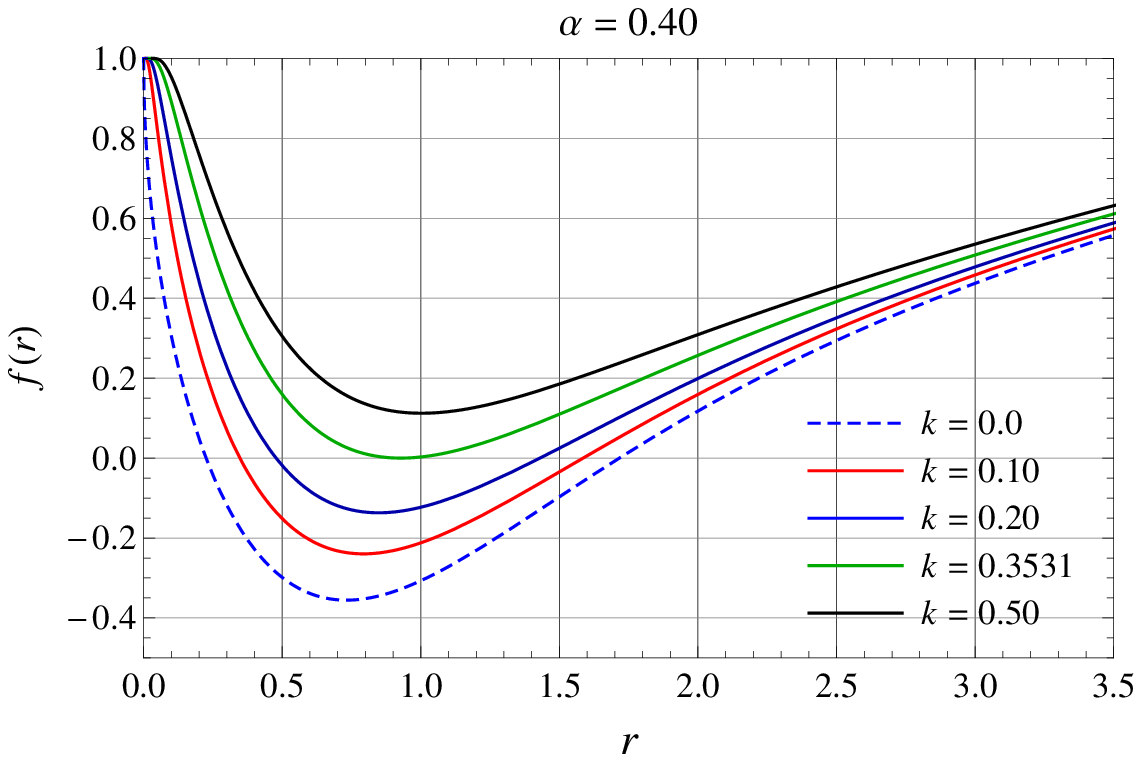}
	\end{tabular}
	\caption{\label{fig:th} Plot of metric function $f(r)$ vs $r$  for different values of  deviation parameter $k$ with GB coupling constant   $\alpha$ = 0.2 and 0.4. }
	\label{f1}
\end{figure*}

The solution (\ref{sol1}) can be  characterized by the mass $M$,  the cosmological constant $1/l^2$, the GB coupling constant  $\alpha$, and  the deviation parameter $k$, which is assumed to be positive. For definiteness we call the solution (\ref{sol1}) as the NED charged $4D$ EGB  black hole. 
In general, most of the regular or nonsingular black holes coupled to NED admit two horizons \cite{Ghosh:2018bxg,Kumar:2020xvu,Kumar:2020uyz,Simpson:2019mud,Culetu:2015cna}. 
The horizon radii are  zeros of $g^{rr}=0$ of $f(r_{h})=0$,  which  implies that 
  \begin{equation}
\frac{r_{h}^4}{l^2}+ r_{h}^2-2M r_{h} \exp\left[\frac{-k}{r_{h}}\right]+\alpha  =0. 
  \end{equation}
We solve the above equation numerically to find that it admits multiple roots (cf. Fig.~\ref{fig:th}) and depending on the choice of parameters $M,\alpha$, and $k$, we have two distinct positive roots $r_{h}=r_{\pm}$ corresponding to two horizons, namely, the smaller inner or Cauchy horizon ($r_{-}$), and the outer  event horizon ($r_{+}  > r_{-}$). It  turns out that, for a given $M$ and $\alpha$ there exists  a critical $k_E$ and radius $r_{E}$ such that, for $ r_h > 0 $, $ f(r_h)=0 $  admits one double zero at $r_{E}$ if $k=k_E$, and two simple zeros at $r_{\pm}$ if $k< k_E$ (cf. Fig.~\ref{fig:th}). These two cases therefore describe, respectively, a regular extremal black hole with degenerate Killing horizon, and a regular non-extremal black hole with both outer and inner Killing horizons.  Whereas if  $k>k_E$, $f(r)=0$ has no roots or a black hole does not exist.  Using similar arguments, we can also determine $\alpha_E.$. It is evident from Fig. \ref{f1} that the event horizon radius $r_{+}$ decreases with increase in the deviation parameter ($k$) and the GB coupling constant  $\alpha$.
\begin{figure*} [b!]
	\begin{tabular}{c c c c}
		\includegraphics[width=0.55\linewidth]{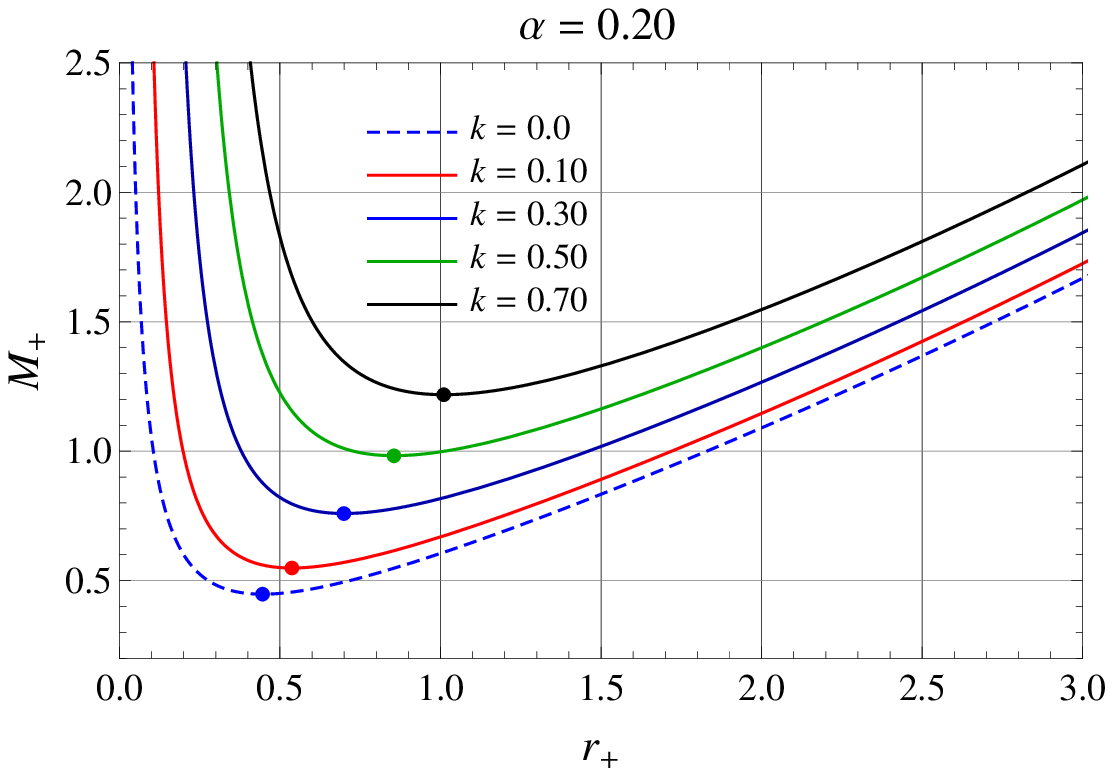}&
\hspace*{-0.8cm}			       
        \includegraphics[width=0.55\linewidth]{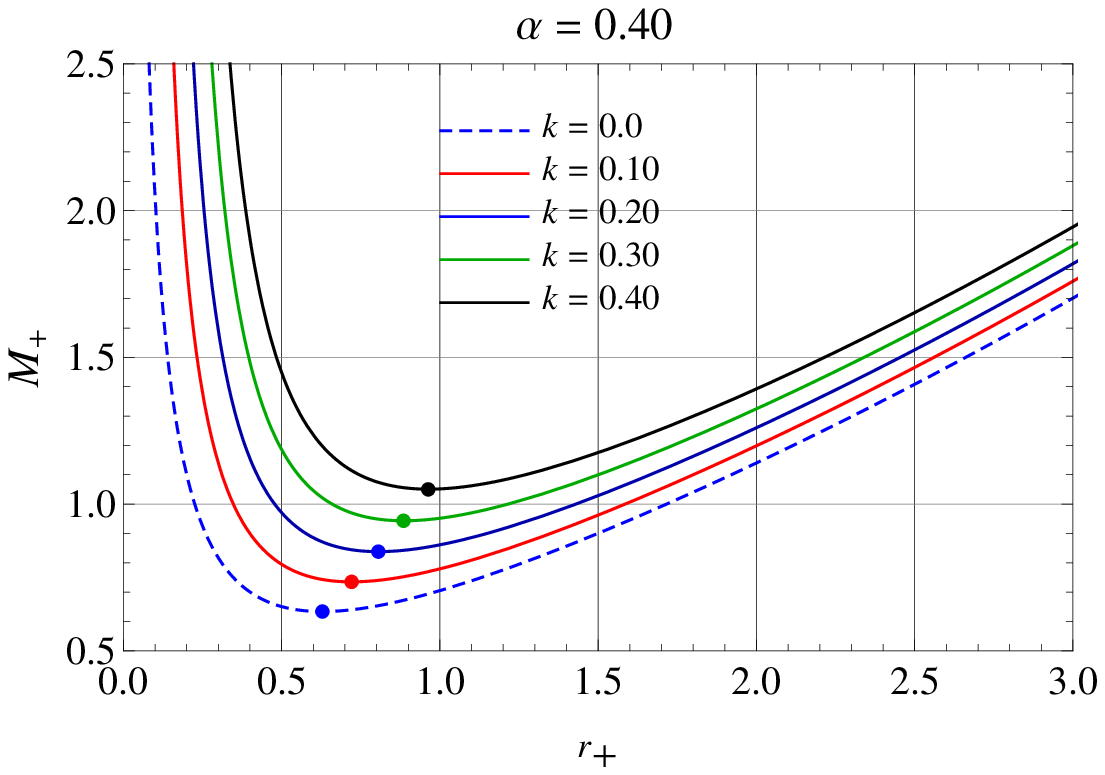}
	\end{tabular}
	\caption{Plot of mass $M_+$  vs horizon radius $r_+$ for different values of GB coupling constant  $\alpha$ and deviation parameter $k$. Colored points correspond to the minimum mass $ M_{+}^{\text{min}}$ and degenerate horizon radii $r_E$ for the black hole existence.}
	\label{mass}
\end{figure*}
\section{Black hole thermodynamics}\label{sect3}
It may be useful to investigate how the NED affects  the thermodynamical properties of   $4D$ EGB black holes; therefore we calculate the  thermodynamic quantities associated with the  NED charged $4D$ EGB AdS  black hole.  We note that the gravitational mass of a black hole, determined by $f(r_+)=0$ \cite{Ghosh:2014pga}, which reads as
\begin{eqnarray}
M_+=\frac{r_{+} \exp{(k/r_{+})}}{2}\left(1 +\frac{\alpha}{r_{+}^2}+\frac{r_+^2}{l^2}\right),
\label{mass1}
\end{eqnarray}
and for $k=0$ it reduces to that for the  $4D$  EGB AdS black hole \cite{gla},  to the  $4D$  charged EGB black hole \cite{Fernandes:2020rpa} for $r\gg k$, and for $\alpha\rightarrow 0$  we obtain  the mass for the regular black hole  in AdS spacetime.   
Let us analyze the effects of NED on the $M_{+}$, which is depicted in the Fig.~\ref{mass}, where we have shown  $M_{+}  $  as a function of horizon radius $r_+$ for different values of $k$ and $\alpha$, and compared also with the neutral ($k=0$) $4D$ EGB counterpart. A minimum  mass $ M_{+}^{\text{min}} $  for the existence of $4D$ EGB  black holes (charged or neutral) occurs at relatively smaller radii $r_{+}$ and also $r_{+}^{\text{min}}$, where the minimum mass appears, are larger in the NED charged case ($k\neq 0$) and so is the  $ M_{+}^{\text{min}} $ values when compared with  the neutral ($k=0$) $4D$ EGB counterpart (cf. Fig.~ \ref{mass}) \cite{gla}. Figure \ref{mass} infers that for $M= M_{+}^{\text{min}}$, the black hole possesses a degenerate horizon at $r_{+}^{\text{min}}$, whereas, for $M>  M_{+}^{\text{min}}$ two distinct horizons exists; nevertheless, $r_{+}^{\text{min}}$ can be identified as $r_E$. Hence the NED charged $4D$ EGB extremal black hole are heavier when compared to the neutral ones. 

The Hawking temperature \cite{Hawking:1974sw} can also help us to  understand  the NED effects on the final stage of the  black hole evaporation. The Hawking temperature of a black  hole can be obtained using the relation $T=\kappa/2\pi$, where $\kappa$ is the surface gravity given  by \cite{Hawking:1974sw,thermo}
\begin{equation} 
\kappa=\left(-\frac{1}{2}\nabla_{\mu}\xi_{\nu}\nabla^{\mu}\xi^{\nu}\right)^{1/2},
\end{equation}
and $\xi^{\mu}=\partial/\partial t$ is the timelike  Killing vector for  the black hole metric (\ref{metric}). 
On using the our solution (\ref{sol1}), the Hawking temperature for the NED charged $4D$ EGB black hole reads 
\begin{equation}\label{temp1}
T_+  =\frac{1}{4\pi r_+^2(r_+^2+2\alpha)}\left[\frac{r_+^4(3r_+-k)}{l^2}+r_+^2(r_+-k)-\alpha(r_++k)\right].
\end{equation}
The temperature of the NED charged $4D$ EGB black hole  (\ref{temp1}) reduces to that of the $4D$ charged EGB black hole \cite{Fernandes:2020rpa} when $r\gg k$, $4D$  EGB black hole \cite{gla} in the limit of $k=0$,  $4D$ AdS regular black hole  when $\alpha \to 0$, and  also to the  Schwarzschild black hole for  $\alpha \to 0$ and $ k = 0$.

\begin{figure*}[b!]
	\begin{tabular}{c c}
\includegraphics[width=0.54\linewidth]{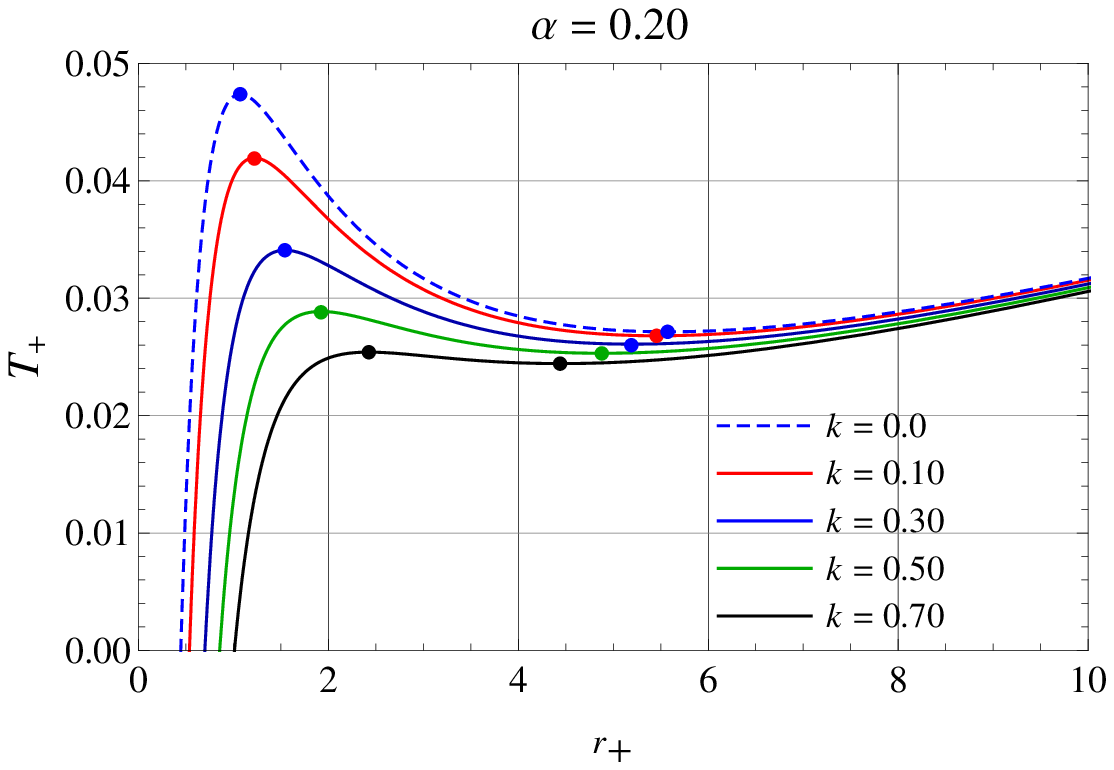}&
\hspace{-0.8cm}
\includegraphics[width=0.54\linewidth]{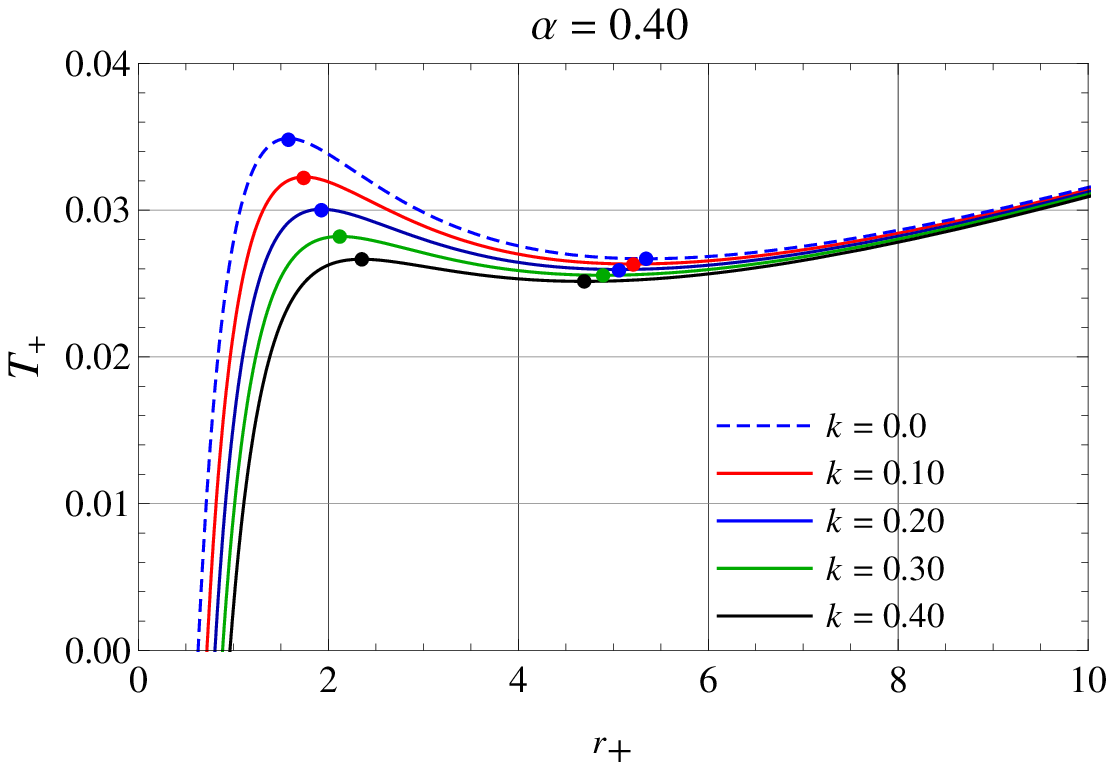}
	\end{tabular}
	\caption{\label{fig:th1} Plot of Hawking temperature $T_+$  vs horizon radius $r_+$  for different values of GB coupling constant  $\alpha$ and deviation parameter $k$. Colored points correspond to local minima and maxima of temperature.}
\end{figure*}
We plot  the  NED charged $4D$ EGB  black hole Hawking temperature ($T_+$) as a function of horizon radii ($r_{+}$) for different values of $k$ and fixed $\alpha$ (cf. Fig. \ref{fig:th1}). From  Fig.~\ref{fig:th1}, it is evident that with  decreasing $r_+$ the Hawking  temperature decreases to attain  a local minimum $T_{\text{min}}$ at horizon radii $r_+^b$ and then grows to a local maximum $T_{\text{max}}$ at radii $r_+^a$ ($r_+^a< r_+^b$), then further steeply drops to zero temperature at the critical radius $r_{+}^{\text{min}}$: Colored points in Fig.~\ref{fig:th1} represent the $T_{\text{max}}$ and $T_{\text{min}}$ at radii $r_+^a$ and  $r_+^b$. It turns out that the local maximum and minimum values of the Hawking temperature decreases with increase in the values of NED parameter $k$ as well as with GB coupling constant  $\alpha$ (cf. Fig. \ref{fig:th1}). The temperature $T_{\text{max}}$ for the NED charged $4D$ EGB  black hole is lower when compared with analogous $4D$ EGB case. We note that for the extremal black holes with the degenerate horizon radius $r_{E}$ (or $r_{+}^{\text{min}}$) the temperature vanishes, $T_+=0$, thereby in the end stage of Hawking evaporation we are left with the zero temperature extremal black holes with minimal mass as a stable remnant. It is evident from Fig.~\ref{fig:th1}, that the NED charged $4D$ EGB  black holes have larger remnant size compared to the uncharged case. Furthermore, for $ T_{\text{min}}\leq T_+\leq T_{\text{max}}$ there exists two black hole configurations  at equilibrium, such that the smaller black hole, represented by the branch with $r_+< r_+^b$ is thermodynamically unstable, whereas the larger black hole with $r_+> r_+^b$ is thermodynamically stable.

We can consider the black hole as a canonical ensemble system where the chemical potential $\phi$, associated with the NED charge $q$, is held fixed.  We calculate the entropy ($S_{+}$) of the black hole in terms of the horizon radius $r_+$. In GR the entropy satisfies the black hole's area law $S_{+} = A/4$.  The black hole behaves as a thermodynamic system; quantities associated with it must obey the first law of thermodynamics $dM_+ = T_+dS_+ \,+\, \phi_+ dq$. The entropy \cite{thermo}, for constant charge $q$,  can be obtained by  integrating the first law as
\begin{eqnarray}\
S_+=\frac{A}{4}\left[\left(1+\frac{k}{r_+}\right)\exp\left[\frac{k}{r_+}\right]-\frac{k^2+4\alpha}{r_+^2}\text{Ei}\left[\frac{k}{r_+}\right]\right],
\label{entropy}
\end{eqnarray}
where $\text{Ei}\left[k/r_+\right]$ is the exponential integral function. Black holes, as thermodynamic systems, should have positive entropy and the cases when it becomes negative will be excluded from our discussion. The entropy of the $4\-D$ EGB black hole, in the absence of NED charge ($k=0$),  becomes 
\begin{eqnarray}
	S_+=\frac{A}{4}+2\pi \alpha \log \left[\frac{A}{A_0}\right], 
	\label{entropy1}
\end{eqnarray}
with $A= 4 \pi r_+^2$ and $A_0$ is a constant \cite{Fernandes:2020nbq,Singh:2020nwo}. 
This is the common area law known as the Bekenstein-Hawking area law for  $4D$ EGB  black holes having correction terms  \cite{Cai:2009ua,Fernandes:2020nbq}.  However, it is interesting to note that the entropy of the black hole has no effect of an AdS background and is same as in the asymptotic flat case. Further, we also notice from Eq.~(\ref{entropy}) that the thermodynamic entropy can become negative due to NED.  
\subsection{Global stability }
The Hawking-Page \cite{hp} phase transition, states that asymptotically AdS Schwarzschild black holes are thermally favoured when their temperatures are sufficiently high. In contrast, the pure AdS background spacetimes are preferred at comparatively low temperatures and there occurs a phase transition between the thermal AdS and the AdS black holes at some critical temperature. The phase transition has been widely studied for higher dimensional EGB gravity asymptotically AdS black holes \cite{egbads,egbnads,Cho:2002hq} and we wish to analyse this for the NED charged $4D$ EGB  black hole via Helmholtz's free energy.  The reason for this is that even if the black hole is locally thermodynamically stable, it could be globally unstable or vice-versa \cite{Herscovich,Cho:2002hq,Cvetic:2001bk}. The Helmholtz  free energy of a black hole can be defined as \cite{Herscovich}
\begin{equation}\label{fe}
F_+ =M_+-T_+S_+,
\end{equation}
which for the NED charged $4D$ EGB  black hole reads as
\begin{align*}
F_+=&\frac{1}{4 l^2 r_+^2 (r_+^2 + 2 \alpha)}\Bigg[2 \exp\Big[\frac{k}{r_+}\Big] r_+ (r_+^2 + 2 \alpha) \Big(r_+^4 + 
l^2 (r_+^2 + \alpha)\Big) + \Big(-3 r_+^5 + l^2 r_+ (-r_+^2 + \alpha) \nonumber\\
&+k \left(r_+^4 + l^2 (r_+^2 + \alpha)\right)\Big)\times\Big(\exp\Big[\frac{k}{r_+}\Big]
r_+ (k + r_+) - (k^2 + 4 \alpha) \text{Ei}\Big[\frac{k}{r_+}\Big]\Big)\Bigg].
\end{align*}
\begin{figure*} [b]
	\begin{tabular}{c c}
		\includegraphics[width=0.55\linewidth]{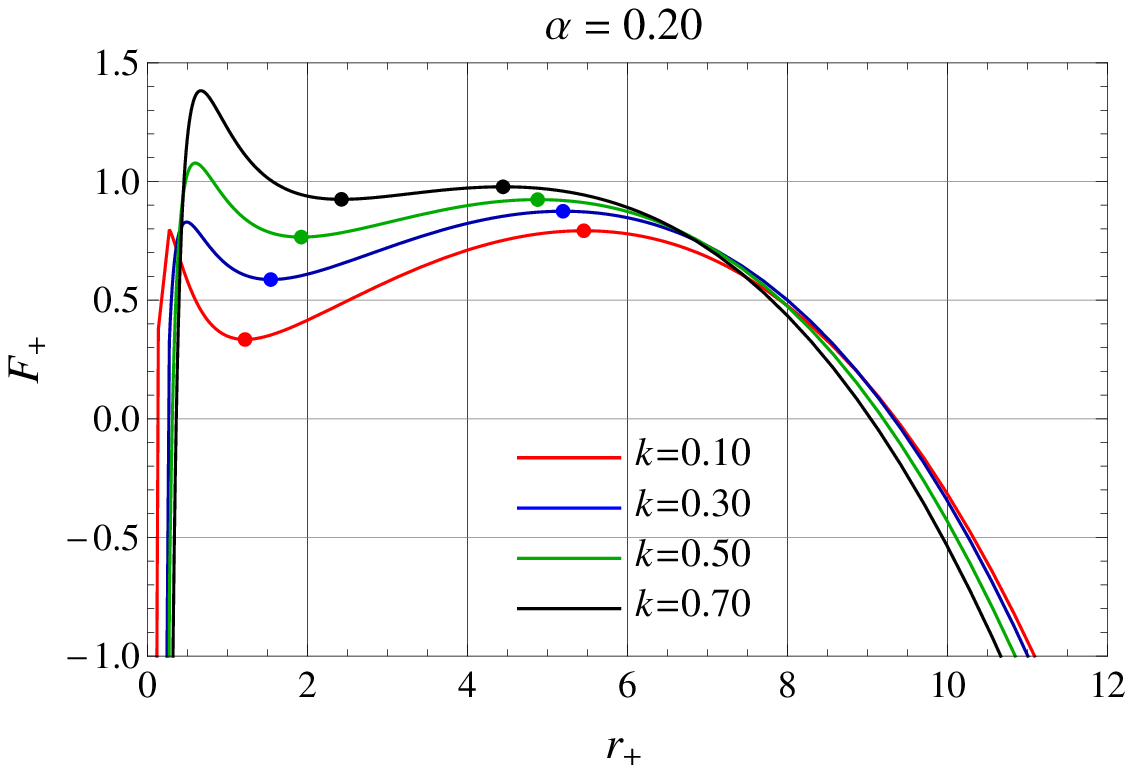}&
		\hspace{-0.9cm}		
		\includegraphics[width=0.55\linewidth]{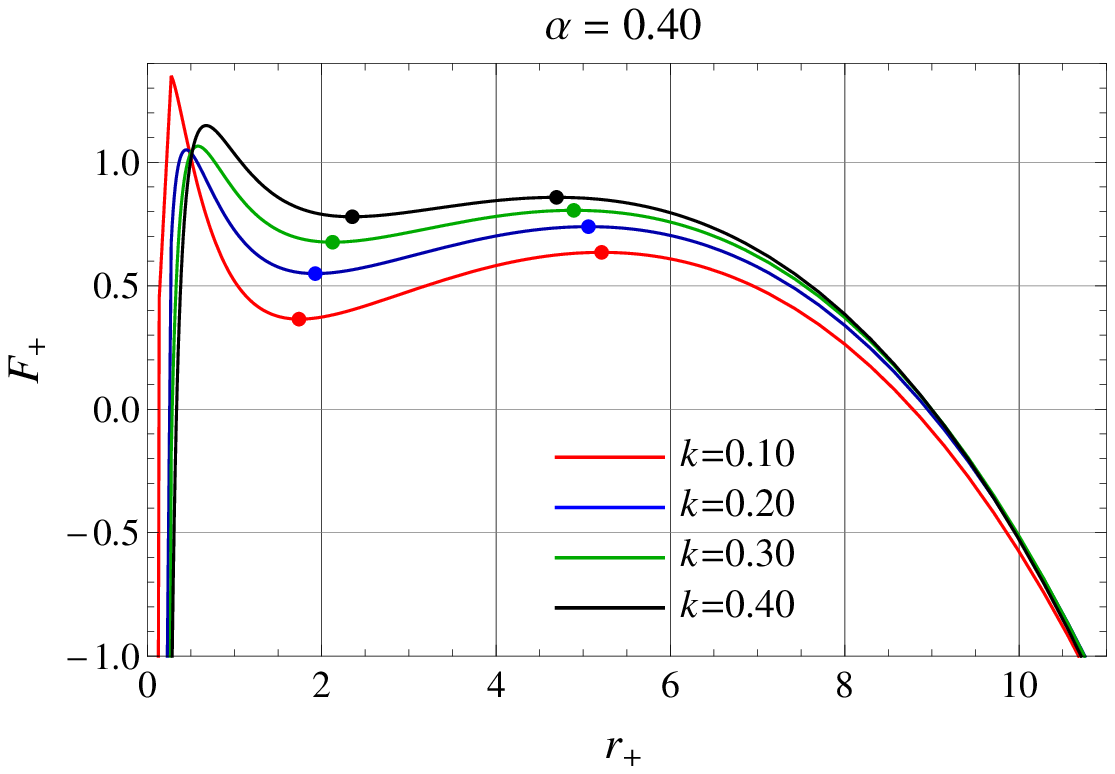}
	\end{tabular}
	\caption{\label{fig:FE} Plot of free energy $F_+$  vs horizon radius $r_+$ for different values of GB coupling constant  $\alpha$ and deviation parameter $k$. Colored points correspond to local minima and maxima of free energy.}
	\label{free1}
\end{figure*}
\begin{figure*}
	\begin{tabular}{c c}
		\includegraphics[width=0.52\linewidth]{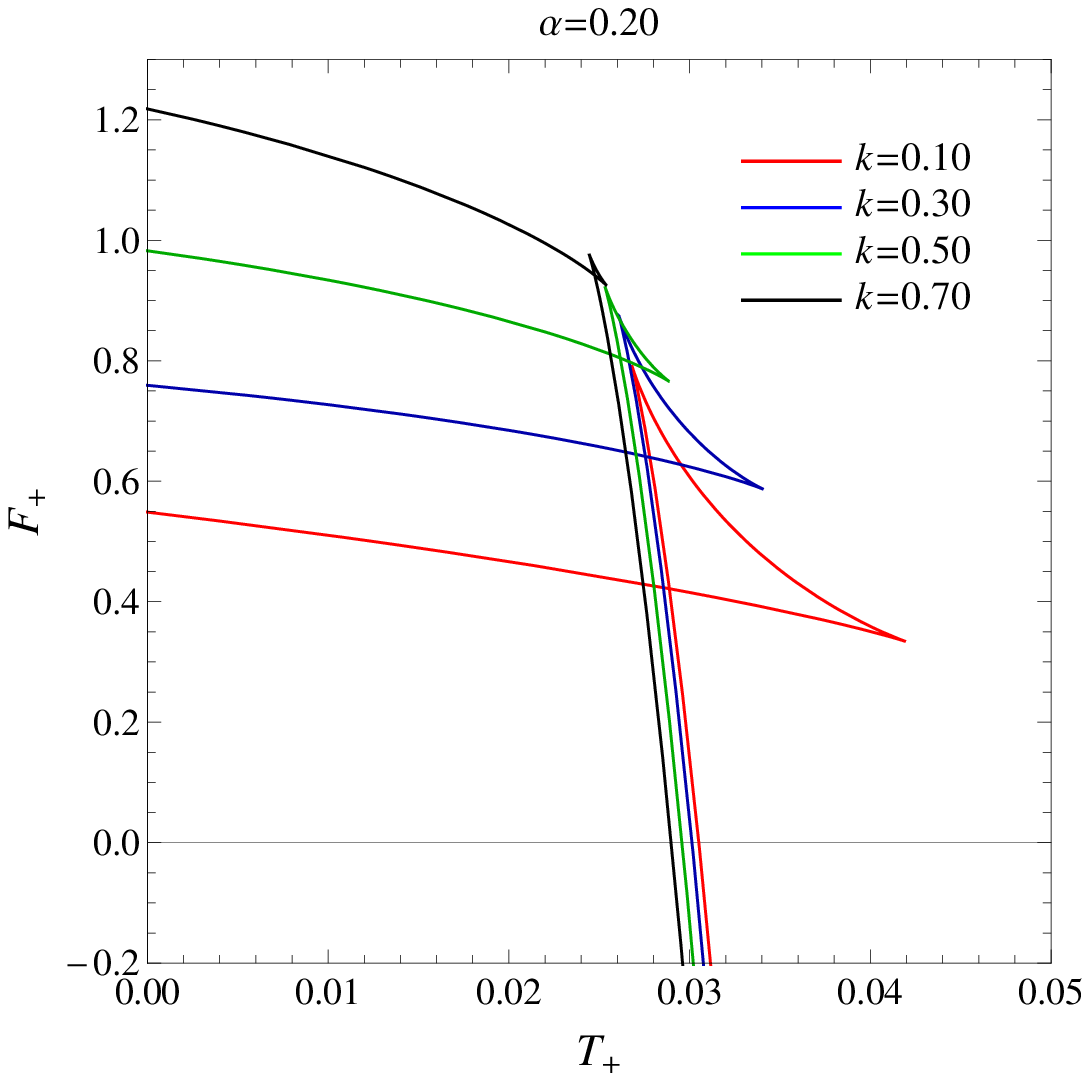}&
		\hspace{-0.4cm}
		\includegraphics[width=0.52\linewidth]{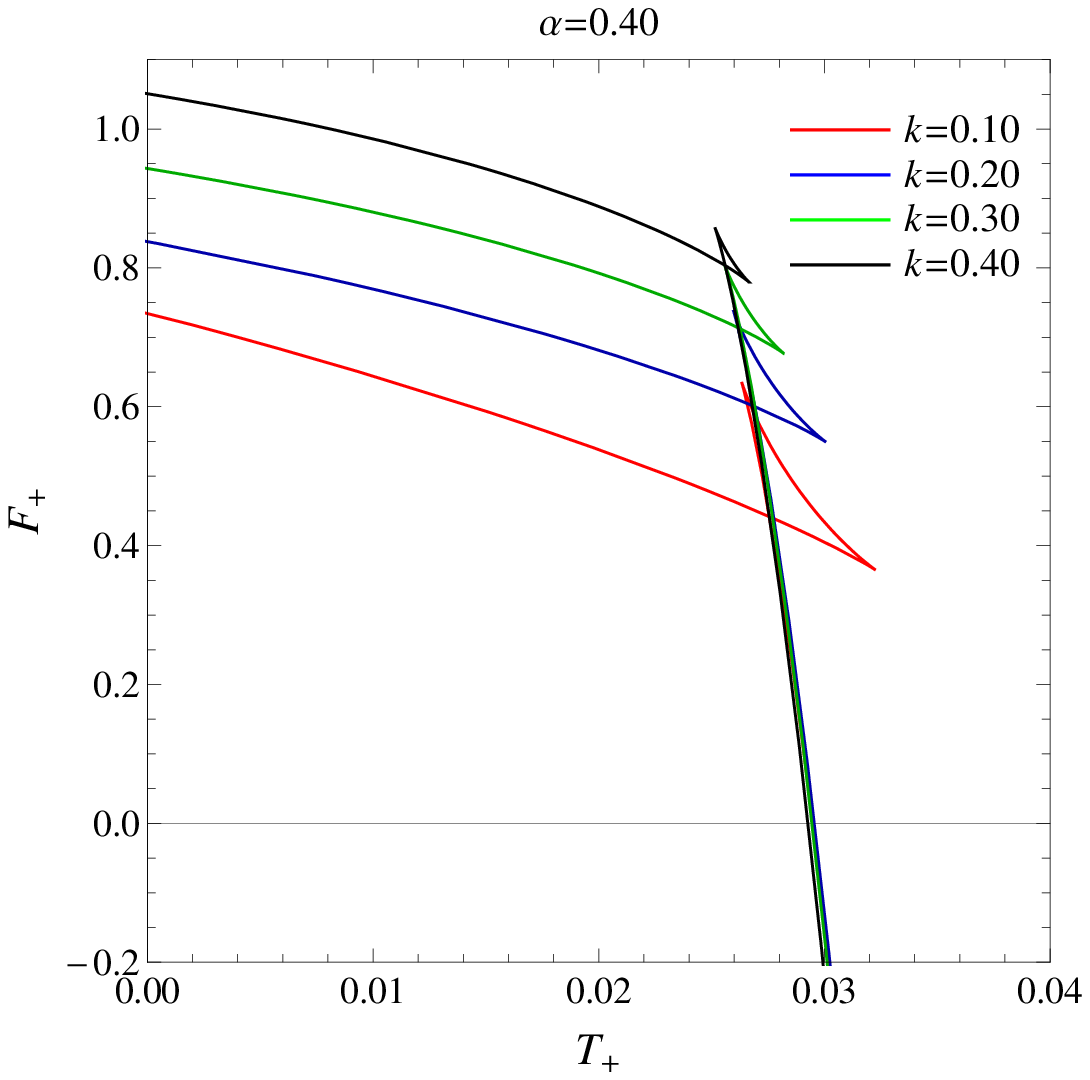}
	\end{tabular}
	\caption{\label{fig:FT} Plot of free energy $F_+$  vs horizon temperature $T_+$  for different values of GB coupling constant  $\alpha$ and deviation parameter $k$.}
\end{figure*}
Generally, it is demonstrated that black holes with negative values of $ F_+ $ are more thermodynamically stable. The Helmholtz free energy of NED charged $4D$ EGB  black holes  for various values  of parameters $k$ and  $\alpha$ is depicted in Fig.~\ref{free1}. As it is shown in  Fig.~\ref{free1}, the free energy $F_{+}$ for various $k$,  have local minimum and local maximum, respectively, at horizon radii $r_{+}^a$ and $r_{+}^b$ with $r_+^b>r_+^a$, which can be identified as the extremal points of the Hawking temperature shown in Fig.~\ref{fig:th1} and where the specific heat capacity $C_{+}$ diverges. For $r_+>r_{+}^b$, the free energy $F_{+}$ is a monotonically decreasing function of $r_{+}$  and becomes negative at large $r_+$, i.e., $F_+(r_{\text{HP}})=0$ such that $F_+>0$ for $r_+<r_{\text{HP}}$ and $F_+<0$ for $r_+>r_{\text{HP}}$. Whereas for $r_+<r_{+}^b$ the free energy $F_+$ decreases with decreasing $r_+$ and attains a local minimum at $r_+^a$ with $F_+(r_+^a)>0$, and with further decreasing $r_+$ the $F_+$ starts increasing (cf. Fig. \ref{free1}). The Hawking-Page first order phase transition occurs at $r_+=r_{\text{HP}}$, where the free energy turns negative viz., $r_{\text{HP}}>r_+^b$ . Thus the larger black holes, with horizon radii $r_+>r_{\text{HP}}$, are thermodynamically globally stable. However, at very small horizon radii the Hawking temperature is negative and hence not physical for global stability. This is exactly in accordance with the Hawking-Page phase transition in GR \cite{hp}. In addition, one can find the temperature $T_+=T_{\text{HP}}$ at which Hawking-Page phase transition happens, in terms of the horizon radius $r_+$, by solving $F(T_{\text{HP}})=0$. One can notice that $T_{\text{HP}}> T_{\text{min}}$ (cf. Fig.~\ref{fig:FT}).  Therefore for $T_+>T_{\text{HP}}$ we find that the black hole solution is favored globally with respect to the thermal AdS background solution as $F_+<0$. While for $T_{\text{min}}<T_+<T_{\text{HP}}$ the radiation in the AdS background solution is globally favoured over the black hole as $F_+>0$. In particular, for $\alpha=0.20$, and $k=0.10, 0.30, 0.50$, and $0.70$, the horizon radii $r_+^a$ are, respectively, $r_+^a=1.22133, 1.54263, 1.92382, 2.42639$ where $T_+$ admits a local maximum and $F_+$ attains a  local minimum, and $r_+^b$ are, respectively, $r_+^b=5.45688, 5.1966, 4.87942, 4.44678$ where $T_+$ admits a local minimum and $F_+$ attains a local maximum. Whereas the horizon radii $r_{\text{HP}}$, for $\alpha=0.20$, and $k=0.10, 0.30, 0.50$, and $0.70$ are, respectively, $r_{\text{HP}}=9.35535, 9.33188, 9.19912, 9.04365$ and the corresponding temperature are, respectively, $T_{\text{HP}}=0.0305108, 0.0301342, 0.0295824, 0.0289863$. Therefore one can infer that, with increasing $k$, the horizon radii $r_{\text{HP}}$ and the corresponding Hawking temperature $T_{\text{HP}}$ for the Hawking-Page phase transition decrease. In Fig.~\ref{fig:FT}, we have shown the behaviour of $F_+$ with $T_+$. The behaviour at small $r_{+}$ is bit unusual from GR and higher dimensional EGB which is attributed to the logarithmic correction term.

\subsection{Local stability}
\begin{figure*}[h!]
\begin{tabular}{c c c c}
\includegraphics[width=0.54\linewidth]{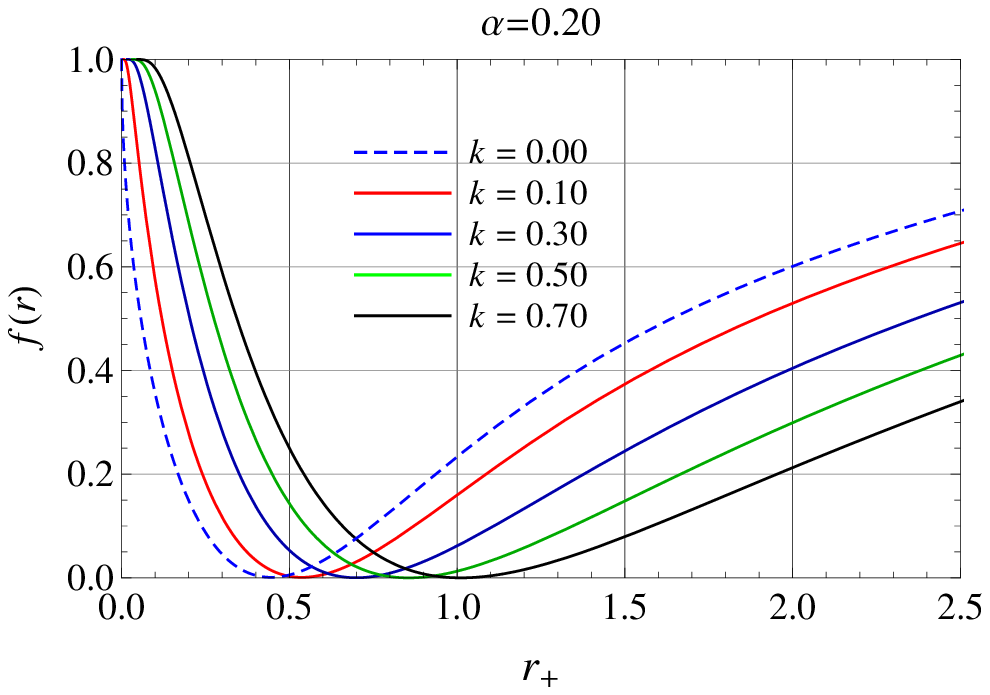}
\hspace{-0.85cm}
\includegraphics[width=0.54\linewidth]{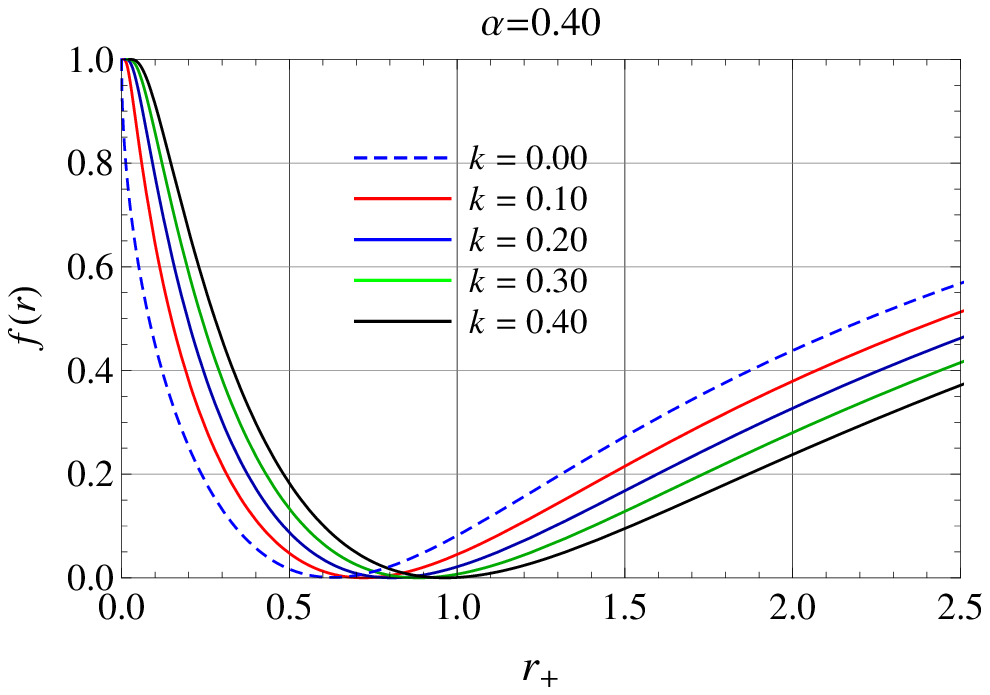}
\end{tabular}
\caption{The plot of metric function $f(r)$ as the function horizon radius $r_+$ for  different values of deviation parameters $k$  and $\alpha$.}
\label{rem1}
\end{figure*}

Having discussed  the conditions for global thermodynamical stability of a NED charged $4D$ EGB  black hole, we turn our attention to the local thermodynamics by computing the heat capacity which  informs us about the thermal stability of the black hole under temperature fluctuations. The reason to consider local stability is that even when a black hole configuration is globally stable, it can be locally unstable \cite{hr}. We use the canonical ensemble where the charge is fixed to investigate local instability. The local thermodynamical stability of the black hole depends on behaviour of the heat capacity $C_+$; positive specific heat, $C_+>0$, infers that the black hole is locally thermodynamical stable and $C_+<0$ means it is unstable. To  analyse the thermodynamic stability of the NED charged $4D$ EGB  black hole,   we calculate  its heat capacity $C_+$, and check how the NED  affects  thermodynamic stability. The heat capacity of the black hole is given \cite{cai,Ghosh:2014pga}
\begin{eqnarray}\label{SH}
C_+&=&\frac{\partial{M_+}}{\partial{T_+}}=\left(\frac{\partial{M_+}}{\partial{r_+}}\right)\left(\frac{\partial{r_+}}{\partial{T_+}}\right).
\label{sp1}
\end{eqnarray}
Substituting the values of mass and temperature from  Eqs.~(\ref{mass1}) and (\ref{temp1}) in Eq. (\ref{sp1}), we obtain the  heat capacity of  the $4D$ AdS regular EGB  black hole as
\begin{equation}
C_+=-\frac{2\pi\exp\left[\frac{k}{r_+}\right](r_+^2+2\alpha)^2\Big(3r_+^5+{l^2}{r_+}(r_+^2-\alpha)-{k}\left(r_+^4+l^2(r_+^2+\alpha)\right)\Big)}{2k\Big(2r_+^4\alpha-l^2(r_+^4+2r_+^2\alpha+2\alpha^2)\Big)+l^2r_+(r_+^4-5r_+^2\alpha-2\alpha^2)-3r_+^5(r_+^2+6\alpha)}.
\label{28}
\end{equation}
\begin{figure*}
	\begin{center}
	\includegraphics[width=1.09\linewidth]{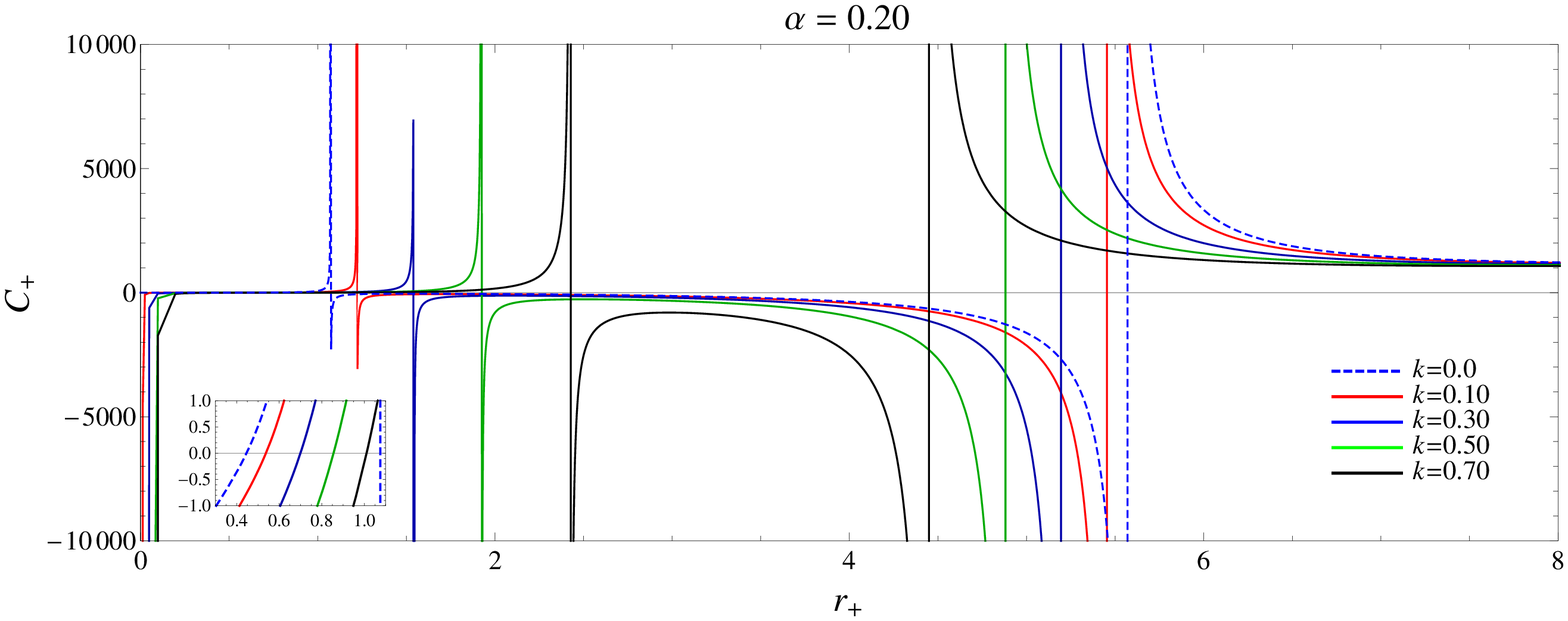}\\
	\includegraphics[width=1.09\linewidth]{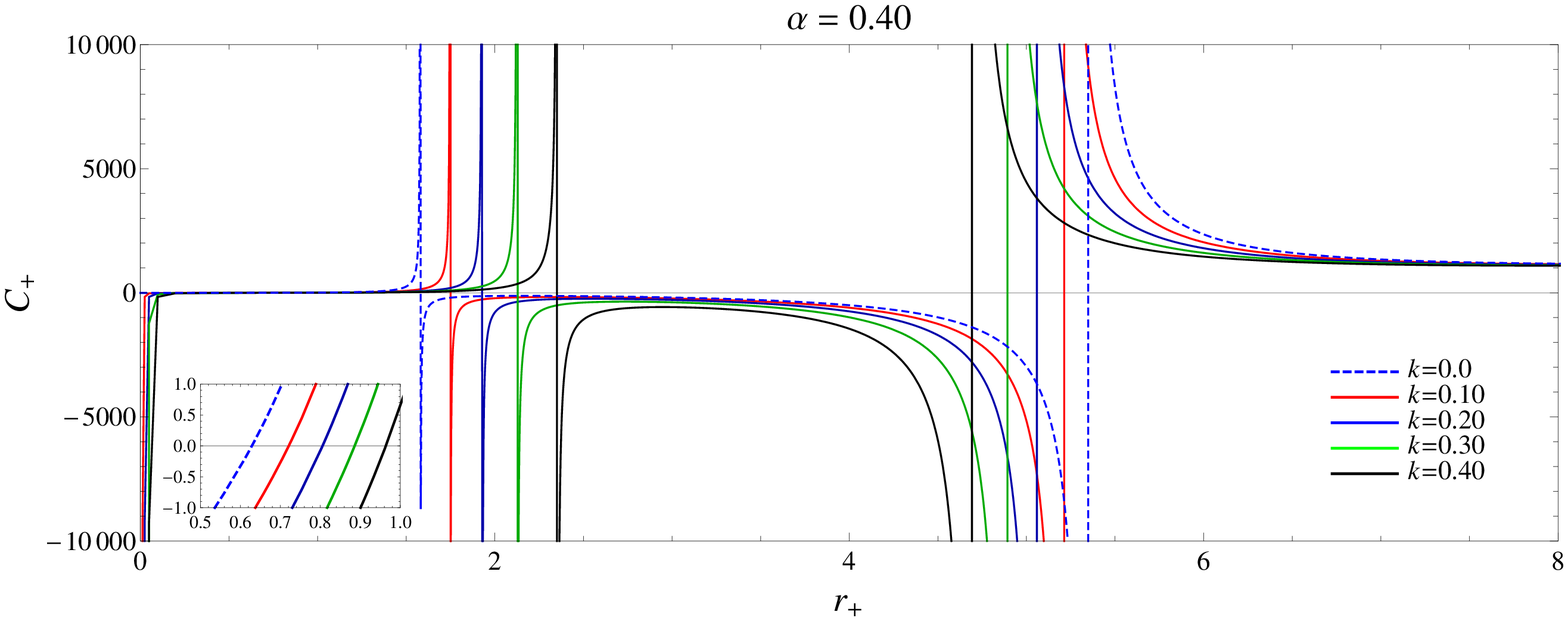}	
	\end{center}\caption{\label{fig:sh} NED Charged $4D$ EGB  black hole specific heat $C_+$ vs. horizon radius $r_+$ for different values of GB coupling parameter $\alpha$ and NED parameter $k$.}
\end{figure*}
To further analyze $C_+$, we plot the  heat capacity in Fig.~\ref{fig:sh} for different values of deviation parameter $k$ and GB coupling constant $\alpha$,  which clearly shows that the heat capacity is discontinuous at the critical radii $r_+^a$ and $r_+^b$ with $r_+^a < r_+^b$  (cf. Fig. \ref{fig:sh}). This signals a second-order phase transition \cite{hp,davis77}. Thus, a NED charged $4D$ EGB  black hole is  thermodynamically stable for  $r_0 <r_+ < r_+^a$ and $r_{+} > r_+^b$, whereas it is thermodynamically unstable for  $r_+^a < r_{+} < r_+^b$ and  $r_+<r_0$. It is evident from  Fig.~\ref{fig:sh},  that  the black hole undergoes a phase transition twice, firstly at $r_+^a$ from the smaller stable black hole to larger unstable black holes and then secondly at $r_+^b$ from the smaller unstable black hole ($r_+^a < r_{+} < r_+^b$) to larger stable black holes ($r_{+}> r_+^b$). For fixed value of $\alpha$ and increasing NED parameter $k$, the critical radii $r_+^a$ increase, whereas radii $r_+^b$ decrease (cf. Fig. \ref{fig:sh}). The heat capacity (\ref{28}), in the limit $\alpha\rightarrow0$, reduces to 
 the value for the analogous GR case, which is the heat capacity of the asymptotically $4D$ regular black holes. 
The heat capacity Eq.~(\ref{28}), in the absence of deviation parameter ($k=0$), reduces for the $4D$ AdS EGB  black hole  \cite{Singh:2020nwo}. \textbf{ In contrary, the Hayward \cite{Kumar:2020xvu} and Born-Infeld \cite{Yang:2020jno} $4D$ EGB black holes undergo phase transition only once. }
\paragraph{Black hole remnant:}
We finally  comment on the black hole remnant  which  is  considered as a source for dark energy \cite{jh}  and also serves as one of the potential candidates to resolve the information loss puzzle \cite{jp}. It turns out that  the extremal black hole with degenerate horizon is given by 
\begin{equation}
f(r_E)= f'(r_E)=0.
\label{fr}
\end{equation}
Thus at the extremal black hole one obtains $r_E = r_-=r_+$,  and the temperature decreases with increasing $r_{-}$  and finally vanishes  leaving a regular double-horizon remnant with $M=M_{+}^{\text{min}} $ which is depicted in the   Fig.~\ref{rem1}.  It means that at the late stage of Hawking evaporation the black hole attains a maximum temperature and then cools down at $r_{E}$ with a stable remanent mass $ M_{+}^{\text{min}} $. Thus the  NED charged $4D$ EGB  black hole shrinks to a dS-like core  with a corresponding  remnant of mass $ M_{+}^{\text{min}} $. 
It can be seen that the temperature decreases  with decreasing horizon radius and vanishes when the two horizons coincide $T_+\to 0,\; C_+\to 0$ as $r_- \to r_+$. Hence a NED charged $4D$ EGB  black hole has a zero temperature thermodynamically stable remnant of mass $ M_{+}^{\text{min}} $ and size $r_E$ (cf. Fig. \ref{rem1}).  Thus, we can say that NED charged $4D$ EGB  black hole  has in general two horizons, which degenerate  to one at $M=M_{+}^{\text{min}} $. The black holes has a phase transition where a heat capacity diverges and flips its sign; a mass decreases during evaporation, temperature vanishes at a double horizon thereby sudden halt of evaporation leaving a  double-horizon remnant with  $M=M_{+}^{\text{min}} $. 
\subsection{P-V criticality} \label{sect4}
We are considering  the cosmological constant in the extended phase space, where the pressure is related to $\Lambda$ through $P=-\Lambda/8\pi=3/8\pi l^2$ leads to the interpretation of mass not only as internal energy but also as Enthalpy $H_+$ of thermodynamical system \cite{Kubiznak:2012wp}. This interpretation leads to following relation for the free energy $F_+=H_+-T_+S_+$ of the system \cite{Ong:2016jrh,Chamblin:1999tk}. 
\begin{figure*}
	\begin{tabular}{c c c c}
		\includegraphics[width=0.5\linewidth]{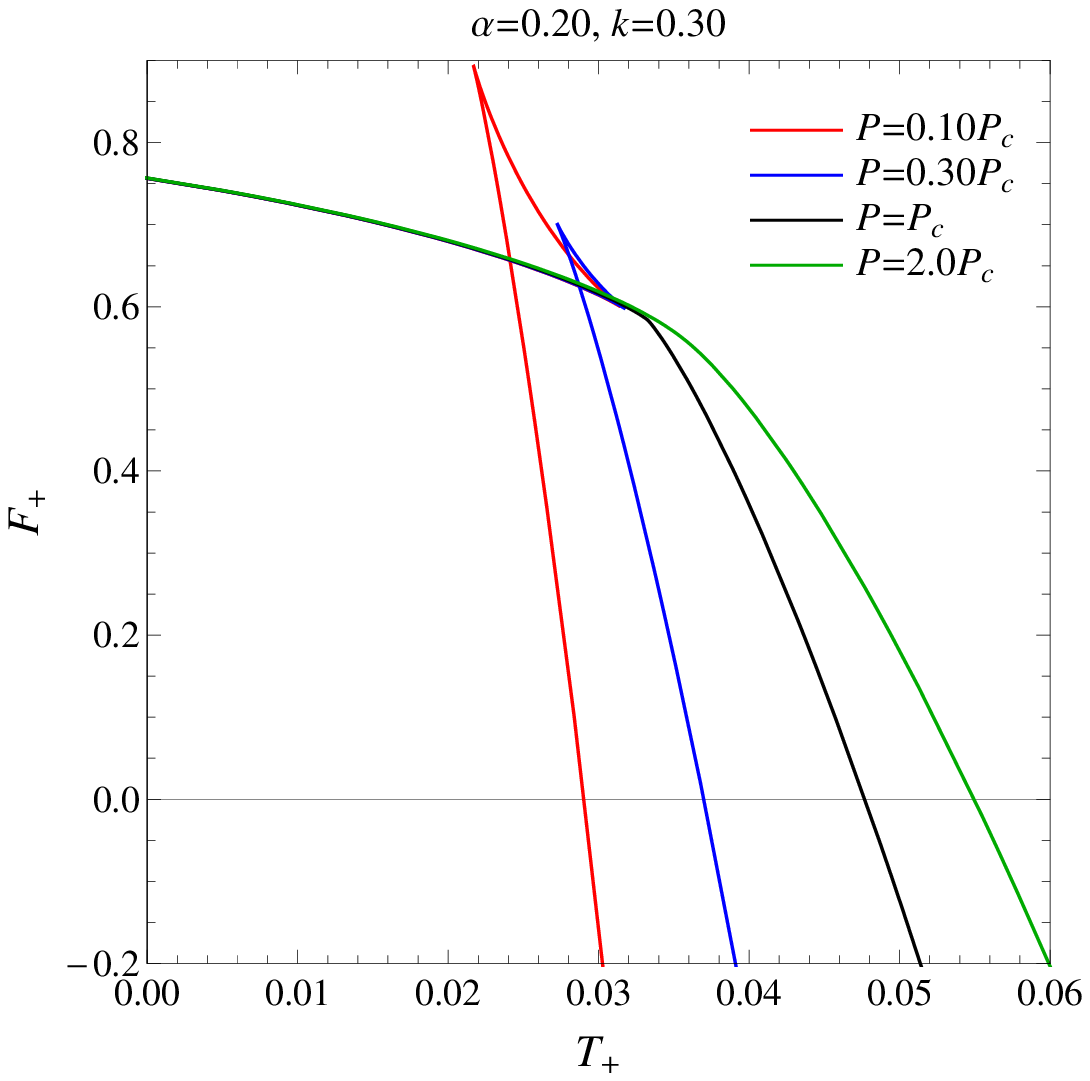}
		\includegraphics[width=0.5\linewidth]{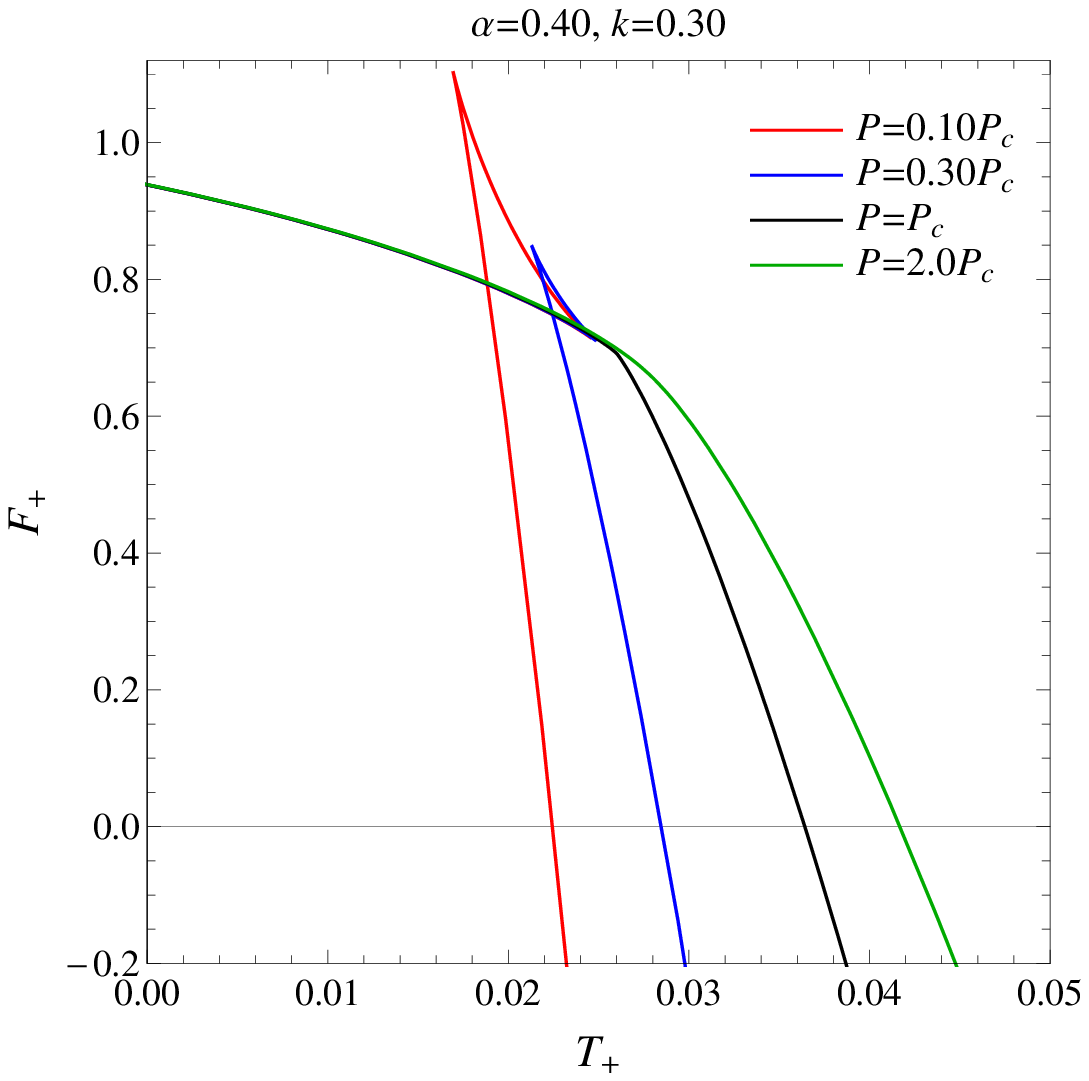}
	\end{tabular}
	\caption{Plot of free energy $F_+$  vs temperature $T_+$ for different values of pressure $P$. The value of critical pressure is $P_c=0.000197371$ for $\alpha=0.20$ and  $P_c=0.0000711748$ for $\alpha=0.40$.}
	\label{sh3}
\end{figure*}
The behaviour of free energy $F_+$ vs temperature $T_+$ for different values of pressure $P$ and GB coupling $\alpha$ is depicted in the Fig. \ref{sh3}. As shown in the Fig. \ref{sh3}, Gibbs Free  develops swallow tail structure when the pressure is below than the critical pressure $P_c$, which infers the first order phase transition. When $P=P_c$, the shallow tail disappear corresponding to the critical point, and when the thermodynamic pressure is larger than the critical pressure $P_c$, no phase transition will occur. 
Using the temperature $T_+$ and specific volume $v=2r_+$ \cite{Cvetic:2010jb}, we can obtain the following equation of state from Eq.~(\ref{temp1})
\begin{eqnarray}
P_+=\frac{3(2\alpha+r_+^2)T_+}{2r_+^4(5r_+-k)}+\frac{3\Big(r_+^2(r_+-k)+\alpha(k+r_+)\Big)}{8\pi r_+^6(5r_+-k)}.
\label{pre}
\end{eqnarray}
Now to calculate the critical values, one can use the inflection point properties \cite{Chamblin:1999tk}
\begin{equation}
\left(\frac{\partial P}{\partial r_+}\right)_T=0,\qquad\left(\frac{\partial^2 P}{\partial r_+^2}\right)_T=0. 
\label{in}
\end{equation}
\begin{figure*}
	\begin{tabular}{c c c c}
		\includegraphics[width=0.5\linewidth]{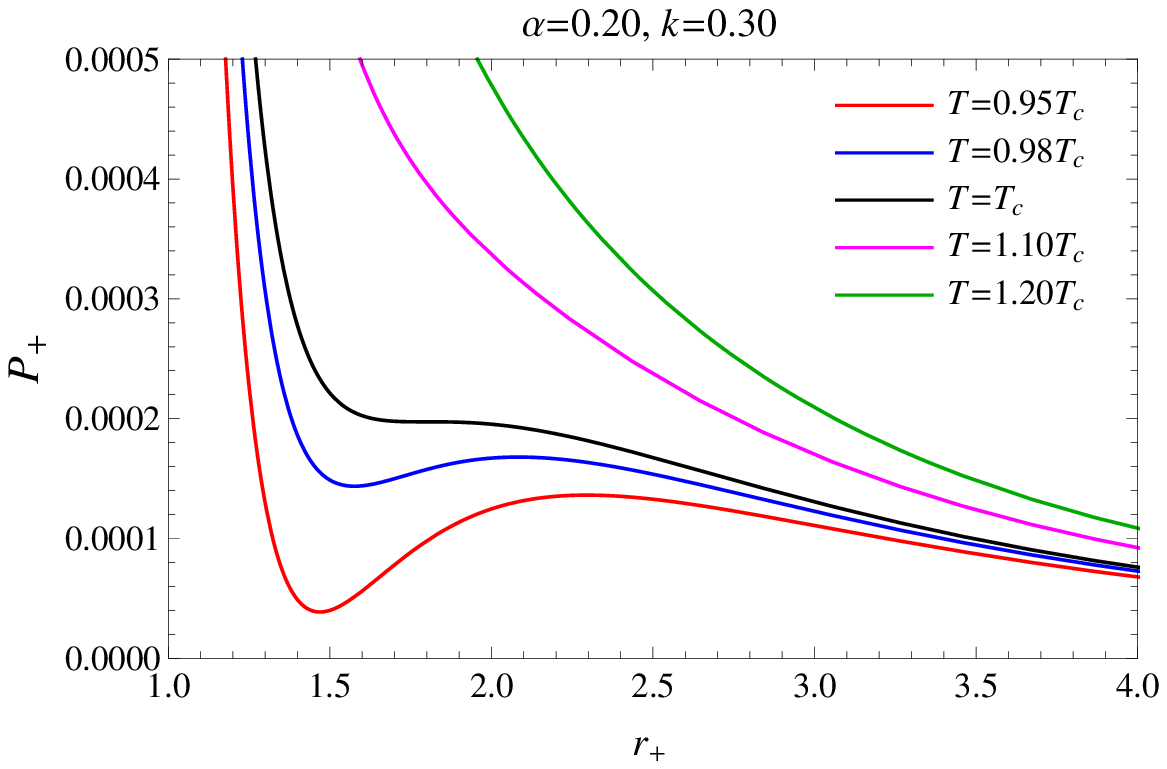}
		\includegraphics[width=0.5\linewidth]{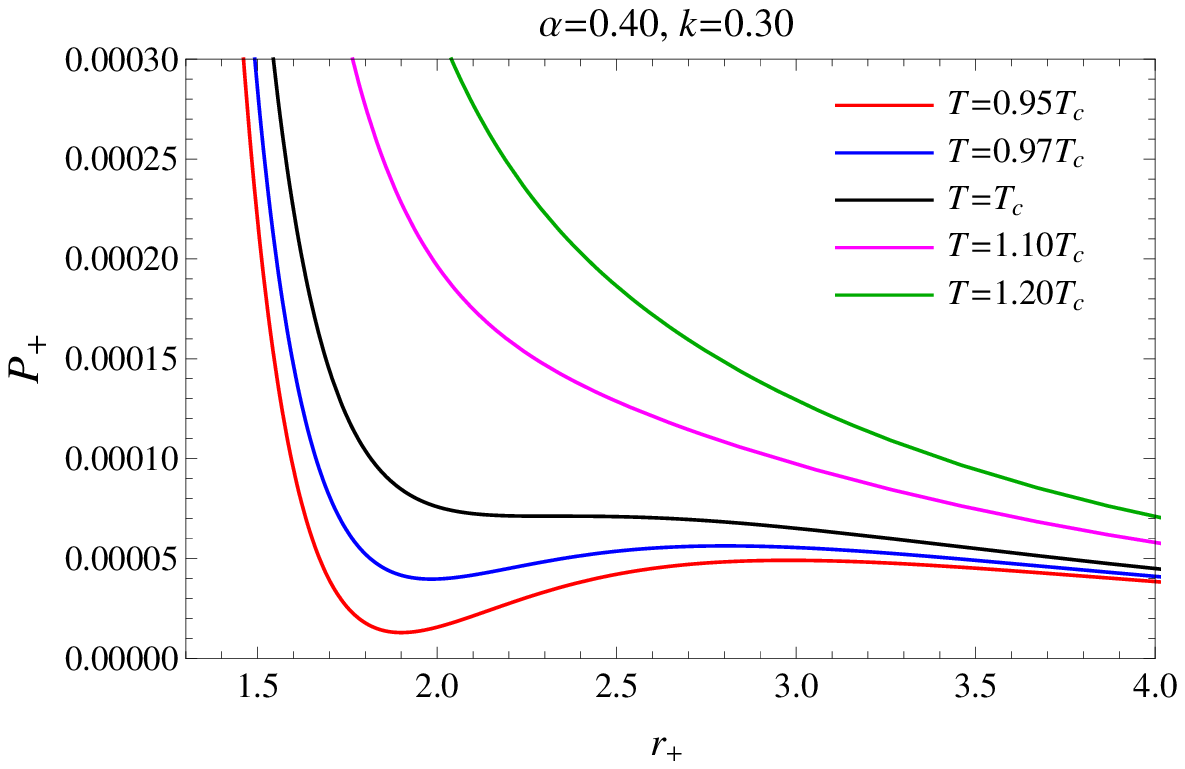}
	\end{tabular}
	\caption{ Plot of pressure $P_+$  vs  radii $r_+$ for different values of temperature $T< T_c,\; T= T_c$ and $T> T_c$. The critical temperature is $T_c=0.0332604$ for $\alpha=0.20$ and  $T_c=0.026007$ for $\alpha=0.40$.  }
	\label{pv5}
\end{figure*}
\begin{center}
	\begin{table*}
		\begin{center}
			\begin{tabular}{p{0.2cm}p{0.2cm} p{1.1cm} p{1.8cm} p{1.8cm} p{1.8cm}  p{0.3cm} p{1.1cm} p{1.8cm} p{1.6cm} p{1.1cm} }
				\hline
				\hline
				\multicolumn{1}{l}{ }&\multicolumn{1}{l}{ }&\multicolumn{1}{l}{ }&\multicolumn{1}{l}{}&\multicolumn{1}{l}{  $\alpha=0.2$}&\multicolumn{1}{l}{ }&\multicolumn{1}{c}{  }&\multicolumn{1}{l}{ }&\multicolumn{1}{l}{ }&\multicolumn{1}{l}{$\alpha=0.4$}\\
				\hline
				\multicolumn{1}{l}{\bf {$k$} } &\multicolumn{1}{c}{ } &\multicolumn{1}{c}{ $r_c$ } & \multicolumn{1}{c}{ $P_c$ }& \multicolumn{1}{c}{$T_c$}& \multicolumn{1}{c}{$P_cv/T_C$} &\multicolumn{1}{c}{}&\multicolumn{1}{c}{$r_c$} &\multicolumn{1}{c}{$ P_c$}   & \multicolumn{1}{c}{$T_c$}& \multicolumn{1}{c}{$P_c v/T_c$} \\
				\hline
				$0.1$ &&  1.46092 & 0.000821072& 0.0452288& 0.0530422&&  1.99844& 0.00201186& 0.0721711& 0.1114181\\
				$0.2$&& 1.62635& 0.000299952& 0.0373109& 0.0261493&& 2.16228& 0.00009706& 0.0283228& 0.0148199\\	
				$0.3$&&  1.79649& 0.000197371& 0.0332604& 0.0213212&& 2.32881& 0.00007117& 0.0260071& 0.0127468\\	
				$0.4$&&  1.97222& 0.000133079&  0.0298641& 0.017577&& 2.4988& 0.00005292& 0.0239742& 0.0110315\\		
				$0.5$ &&  2.15382& 0.000091715& 0.0269926& 0.0146365&& 2.67269& 0.00003984& 0.0221814& 0.0096007\\		
				\hline
				\hline
			\end{tabular}
		\end{center}
		\caption{The numerical value of  critical exponents (radius $r_c$, temperature $T_c$, pressure $P_c$) and global parameter for different values of deviation parameter $k$ and GB Coupling $\alpha=0.2$ and $\alpha=0.40$.}
		\label{tr11}
	\end{table*}
\end{center}

We solved the Eq.~(\ref{in}) for critical horizon radius $r_c$ and critical temperature $T_c$, and used Eq.~(\ref{pre}) to obtain the critical pressure $P_c$. We summarized the numerical values of critical parameters in Table \ref{tr11} for various values of NED charge parameter $k$. In order to elaborate the effect of NED parameter $k$ and GB coupling $\alpha$ we plot the pressure $P_+$ with horizon radius $r_+$ for various isotherms in Fig.~\ref{pv5}. From the Table \ref{tr11}, one can see that for fixed $\alpha$ and increasing $k$, the critical radius $r_c$ increases, whereas critical pressure $P_c$ and critical temperature $T_c$ decrease. In addition the global parameter also decreases with increasing the NED charge $k$.
\section{Conclusion}\label{sect5}
Lately, there has been a surge of interest in {\it regularisation}  of EGB gravity, namely, the limit  $D \to 4 $ of the $D$-dimensional solutions of EGB gravity.  Interestingly, the static spherically symmetric black hole solutions in the various proposed $D \to 4 $  gravities coincide, and incidentally some other theories also admit the identical solution. We have obtained an exact $4D$ static spherically symmetric black hole solution to the $4D$ EGB gravity coupled to the NED which encompasses the black hole solutions of Glavan and Lin \cite{gla} when NED is switched off ($k=0$)  and asymptotically ($r\gg k$)  mimics the charged black holes of Fernandes \cite{Fernandes:2020rpa}. The NED charged $4D$  -AdS black hole metric is characterised by horizons which could be at most two, describing different objects including an extremal black hole with degenerate horizons and non-extremal black holes with two distinct horizons.  The thermodynamic quantities associated with NED charged $4D$  EGB-AdS black hole have been analysed as a function of $r_{+}$, $k$ and $\alpha$. Concerning the thermodynamics properties, we noticed some significant  NED effects and corrections to the previously obtained  $4D$ EGB black holes were discovered. The Hawking temperature, as in the asymptotically flat case, does not diverge as the event horizon shrinks down; instead, it has a local minimum before taking a maximum value for a critical radius and then drops down to zero at degenerate horizons, which happens for larger values of the radius $r_{+}$  due to NED (cf. Fig. \ref{fig:th1})  and also increases with parameter $\alpha$. The entropy of a $4D$ EGB-AdS black hole is exactly the same as in the asymptotically flat case.  The entropy of a black hole in GR obeys the area law, but not for neutral $4D$ EGB-AdS black holes where it has a logarithmic correction term whereas the NED charged $4D$  EGB-AdS black hole is more complicated.  The heat capacity diverges at critical horizon radii $r_+^a$ and $r_+^b$, which depends on the NED parameter $k$, and incidentally, local extrema of the Hawking temperature also occur at these radii. The phase transition is detectable by the divergence of the heat capacity ($C_+$) at  critical radii (changes with  NED parameter $k$), such that the black hole is stable in the region viz:  $r_0 <r_+ < r_+^a$ and $r_{+} > r_+^b$ with positive heat capacity ($C_+>0$), on the other hand the heat capacity is negative ($C_+<0$), when $r_+^a < r_{+} < r_+^b$ and  $r_+<r_0$, indicating the instability of black holes. We also showed that identical black hole solution also exists in the physically motivated alternate regularised $4D$ EGB gravity, based on the Kaluza-Klein-like dimensional reduction procedure. 

We find that NED charged $4D$  EGB-AdS black holes because of exponential mass function leads to a Minkowski-flat core around $ r=0 $ which is in striking contrary with analogous other regular black holes that generally have a de-Sitter core. The simple exponential function in the solution  significantly simplifies the physics in the deep core and is mathematically interesting due to its tractableness.

We find that the NED has a profound influence on the properties of black holes   which may have several astrophysical consequences, for example, on wormholes and accretion onto black holes.  Some of the results presented here are generalizations of previous discussions on  $4D$ EGB  \cite{gla, Fernandes:2020rpa} and GR black holes \cite{Ghosh:2014pga}, to a more general setting.  The possibility of a further generalisation of these results to Lovelock gravity \cite{Konoplya:2020qqh} is an interesting problem for the future. 

\appendix
\section{Solution (\ref{sol1}) by alternate regularization}
One of the potential alternative for the $4D$ regularization of EGB gravity is via the Kaluza-Klein-like reduction of the $D$-dimensional EGB gravity on a $(D-4)$-dimensional maximally symmetric space \cite{Lu:2020iav,Kobayashi:2020wqy}. The resulting theory leads to well defined action principle in $4D$ and describes a scalar-tensor theory of gravity that belongs to a class of Horndeski gravity. Similar procedure was used by Mann and Ross \cite{Mann:1992ar} for obtaining the $D\to2$ limit of Einstein GR. Following \cite{Lu:2020iav}, we start with the $D$-dimensional EGB gravitational action (\ref{action1}) and consider a Kaluza-Klein ansatz
\begin{equation}
ds_D^2=ds_p^2+\exp[2\psi]d\Sigma^2_{D-p},
\end{equation}
where $d\Sigma^2_{D-p}$ is the line element on the internal maximally symmetric space of curvature proportional to $\lambda$, $ds_p^2$ is the $p$-dimensional line element, and $\psi$ is the scalar field depending on the coordinates of external $p$ dimensional space. Redefining the GB coupling as $\alpha\to \alpha/(D-p)$ and taking the limit $D\to p $ in (\ref{action1}), we obtained the $p$-dimensional reduced EGB gravitational action, which for $p=4$ reads as
\begin{align}
\mathcal{I}_4=&\int d^4x\sqrt{-g}\Big[R
-2\Lambda+\alpha\Big(\psi\,\mathcal{L}_{GB}+4G^{\mu\nu}\partial_\mu\psi\partial_\nu\psi-2\lambda R e^{-2\psi} -4(\partial\psi)^2\Box \psi+2\left((\partial\psi)^2\right)^2\nonumber\\
&-12\lambda(\partial\psi)^2e^{-2\psi}-6\lambda^2e^{-4\psi}\Big)-4\mathcal{L}(F)\Big],\label{action2}
\end{align}
and corresponds to the $4D$ regularized EGB gravity action with rescaled GB coupling constant. One can obtain the covariant field equations by varying the action (\ref{action2}) for metric tensor $g_{\mu\nu}$ and scalar field $\psi(r)$ \cite{Hennigar:2020lsl}. To study the static spherically symmetric black hole solution, we consider the metric \textit{ansatz} and scalar field as follows
\begin{equation}
ds_4^2=-\exp[-2\chi(r)]f(r)dt^2+\frac{dr^2}{f(r)}+r^2d\Omega^2_2,\quad \psi=\psi(r).\label{anstaz}
\end{equation}

On substituting the ansatz (\ref{anstaz}) to action $\mathcal{I}_4$ in (\ref{action2}), we obtain the effective Lagrangian  
\begin{align}
L_{\rm eff}=&e^{-\chi}\Big[2(1-\Lambda r^2 - f- r f') +\frac{2}{3}\Big(
3 r^2 f^2 \psi '^3+2 r  \left(-r f'+2 r f \chi '-4 f\right)f \psi '^2-6  \Big(-r f'+2 r f \chi '\nonumber\\
&-f+1\Big)f \psi '-6 (f-1) \left(f'-2 f \chi '\right)\Big)\alpha\psi'+ 4\alpha \lambda e^{-2\psi} \Big(r^2 f' \psi '-2 r^2 f \chi ' \psi '-3 r^2 f \psi '^2+r f'+f-1\Big)\nonumber\\
&-6\alpha\lambda^2 r^2 e^{-4\psi}-4r^2\mathcal{L}(F)\Big]\,.
\end{align}
On using the Euler-Lagrange equations, we obtain the dynamical equations for metric functions $f(r)$ and $\chi(r)$, and scalar field $\psi(r)$. Considering the special case of $\chi(r)=0$ \cite{Lu:2020iav}, these equations for the internally flat spacetime ($\lambda=0$), respectively, read as
\begin{eqnarray}
&&\exp[\psi] \alpha \Big(1 -(1 - r \psi')^2f\Big) (\psi'^2+\psi'')=0,\label{A1}\\
&&\exp[3\psi]\alpha\Bigg[ \Big(2\psi' + (1-r\psi')^2f'\Big)f' -f''- 2(1-r\psi') \left(-2\psi'^2+\psi''-3r\psi'\psi'' \right)f^2  + \Big((1-r\psi')^2f'' + 2\psi''\nonumber\\
&&\qquad\;\;\;\;\;\;\; -2 (-1+r\psi' )f'\left(-3\psi' +2r\psi'^2-r \psi'' 
\right)  \Big)f
\Bigg]=0,\label{A2}\\
&&\exp[3\psi]\Bigg[ 1 -\Lambda r^2  -2r^2\mathcal{L}(F)-(r+2\alpha\psi')f' + \Bigg(-1+\alpha\psi' \Big(-2(1+f)\psi' +r^2f\psi'^3+2\Big(3\nonumber\\
&&+r\psi'(-3+r\psi' ) \Big)f'\Big) + 4\alpha\Big(-1+(-1+r\psi' )^2f\Big)\psi''\Bigg)f 
 \Bigg]=0.\label{A3}
\end{eqnarray}
Solving Eq.~(\ref{A1}), leads to the solution for the scalar field as follow:
\begin{equation}
\psi(r)=\log\Big[\frac{r}{L}\Big] + \log[\cosh(\xi)-\sinh(\xi)], \quad \xi(r)=\int_{1}^{r}\frac{du}{u\sqrt{f(u)}},\label{A4}
\end{equation}
where $L$ is an integration constant. For the scalar field (\ref{A4}), the dynamical equation for $\psi(r)$ in (\ref{A2}) is automatically satisfied, whereas 
using the $4D$ NED Lagrangian density $\mathcal{L}(F)$ from Eq.~(\ref{lf}) and $\Lambda=-3/l^2$, Eq.~(\ref{A3}) yields the solution for metric function $f(r)$ as
\begin{equation}
f_{\pm}(r)=1+\frac{r^2}{2\alpha}\left(1\pm\sqrt{1+4\alpha\left(\frac{2M \exp(-k/r)}{r^3}-\frac{1}{l^2}\right)}\,\right).
\end{equation}

Although this approach of $4D$ regularization of EGB gravity is noteworthy different in spirit from the one proposed by Glavan and Lin \cite{gla}, interestingly, two theories yield exactly same static spherically symmetric black hole solutions. Therefore, the static spherically symmetric NED charged $4D$ EGB-AdS black hole (\ref{sol1}) is indeed an exact solution of $4D$ regularized EGB gravity, irrespective of the followed regularization procedure. However, a larger class of black hole solutions may exists in the $4D$ effective scalar-tensor gravity theory followed by Kaluza-Klein approach \cite{Hennigar:2020lsl,Lu:2020iav,Ma:2020ufk}.

\section*{Acknowledgments}
Authors would like to thank DST INDO-SA bilateral project DST/INT/South Africa/P-06/2016, S.G.G. also thank SERB-DST for the ASEAN project IMRC/AISTDF/CRD/2018/000042. S.D.M. acknowledges
that this work is based upon research supported by the South African Research Chair Initiative of the Department of Science and
Technology and the National Research Foundation. R.K. thanks UGC, Govt. of India for financial support through SRF scheme.


\end{document}